\documentclass{emulateapj}

\newcommand{\mac}{M_{\rm ac}}
\newcommand{\zac}{Z_{\rm ac}}
\newcommand{\yac}{Y_{\rm ac}}
\newcommand{\xac}{X_{\rm ac}}

\newcommand{\msun}{\rm M_\odot}
\newcommand{\rsun}{\rm R_\odot}
\newcommand{\lsun}{\rm L_\odot}
\newcommand{\tsun}{\tau_\odot}
\newcommand{\mearth}{\rm M_\oplus}
\newcommand{\xini}{X_{\rm ini}}
\newcommand{\yini}{Y_{\rm ini}}
\newcommand{\zini}{Z_{\rm ini}}
\newcommand{\xs}{X_{\rm S}}
\newcommand{\ys}{Y_{\rm S}}
\newcommand{\zs}{Z_{\rm S}}
\newcommand{\rcz}{R_{\rm CZ}}
\newcommand{\dc}{\left<\delta c/c\right>}

\newcommand{\taui}{\tau_{\rm ac,i}}

\newcommand{\dtau}{\Delta \tau_{\rm ac}}

\usepackage{graphicx} 
 
\begin{document} 

\title{Solar  models with  accretion. I.  Application to  the  solar abundance
  problem} 
\shorttitle{Solar models with accretion. I.}
 
\author{Aldo M.  Serenelli}
\affil{Instituto de Ciencias del  Espacio (CSIC-IEEC), Facultad de Ciencias,
  Campus UAB, 08193 Bellaterra, Spain} 
\affil{Max  Planck Institute  for  Astrophysics, Karl  Schwarzschild Str.   1,
  Garching, D-85471, Germany} 
  
\author{W. C. Haxton} 
\affil{Department of Physics, MC  7300, University of California, Berkeley, CA
  94720-7300}
\affil{Nuclear  Science  Division,   Lawrence  Berkeley  National  Laboratory,
  Berkeley, CA 94720-8169}

\and

\author{Carlos Pe\~na-Garay}
\affil{Instituto de F\'isica  Corpuscular (CSIC-UVEG), Aptdo.  22085, E-46071,
  Valencia, Spain}

\begin{abstract} 

We  generate new  standard solar  models using  newly analyzed  nuclear fusion
cross  sections and  present  results for  helioseismic  quantities and  solar
neutrino  fluxes. We discuss  the status  of the  solar abundance  problem and
investigate  whether   nonstandard  solar  models  with   accretion  from  the
protoplanetary disk might alleviate the  problem.  We examine a broad range of
possibilities,  analyzing  both  metal-enriched and  metal-depleted  accretion
models and  exploring three scenarios for  the timing of  the accretion.  Only
partial solutions are found: one can  bring either the depth of the convective
zone or the surface helium abundance into agreement with helioseismic results,
but not both simultaneously.  In addition, detailed results for solar neutrino
fluxes show that  neutrinos are a competitive source  of information about the
solar  core  and  can  help  constrain possible  accretion  histories  of  the
Sun. Finally, we briefly discuss  how measurements of solar neutrinos from the
CN-cycle could  shed light on  the interaction between  the early Sun  and its
protoplanetary disk.

\end{abstract}

 
\maketitle 

\section{Introduction} 

Solar models have been very successful in describing the detailed structure of
the    solar    interior    that    emerged    from    helioseismic    studies
\citep{modelS,bp01,turck:2001}.   Further success came  with the  first direct
measurement  of  the  total  $^8$B  neutrino flux  by  the  SNO  Collaboration
\citep{sno:2002},  which proved  to be  in excellent  agreement  with standard
solar model (SSM) predictions \citep{bp01}.  Solar models are sensitive to the
details of their constitutive physics.  Helioseismic predictions, for example,
depend critically  on radiative opacities  and microscopic diffusion,  and are
also  sensitive  to uncertainties  in  input  parameters,  such as  abundances
determined from absorption line analyses of the Sun's atmosphere.  Predictions
of solar neutrino fluxes are sensitive  to the rates of key nuclear reactions.
Solar models have  played a mayor role not only in  understanding the Sun, but
also as a  test bench for constituitive physics of  stellar models in general.
The further development of SSMs -- both improved treatments of the physics and
more accurate  input parameters -- as well  as tests of the  model against new
solar observations are thus of fundamental importance to stellar astrophysics.

Over the  last decade,  the advent of  3D hydrodynamic models  of near-surface
solar  convection \citep{stein:1998,asplund:2000,freytag:2002,wedemeyer:2004},
a more careful selection  of spectral lines (e.g. \citealp{allende:2001}) and,
in some cases, relaxation of the assumption of local thermodynamic equilibrium
in  line  formation  \citep{asplund:2004,bergemann:2010}  have resulted  in  a
thorough reassessment  of the photospheric  solar abundances. The  most recent
and complete  revision by \citet{agss09}, hereafter AGSS09,  shows a reduction
in the  abundances of  the volatile CNO  elements and  Ne ($\sim$0.1-0.15~dex)
with  respect  to  older  compilations of  solar  abundances  \citep[hereafter
  GS98]{gn93,gs98},  as  well  as  smaller  changes  for  abundant  refractory
elements, e.g.  Si and Fe.   One simple figure-of-merit for comparisons is the
photospheric total  metal to  hydrogen fraction, $(Z/X)_\odot$  of $\sim$0.018
for AGSS09 versus $\sim$0.023 for GS98.  The AGSS09 results, however, have not
gone  unchallenged.  Based  on 3D  solar atmosphere  models computed  with the
CO5BOLD    code     \citep{freytag:2002},    \citet[and    references    there
  in]{caffau:2010}  have  derived best-value  solar  CNO  abundances that  are
systematically higher  than those in AGSS09 and  associated uncertainties that
are larger, so that the ranges  are formally consistent with GS98 values.  The
differences  between   the  two  groups  seem   to  arise  not   from  the  3D
hydrodynamical atmosphere models, but from the selection of spectral lines and
assumptions in the line formation calculations.

Nuclear reaction  rates are  another important input  for SSMs.  A  very broad
effort to  critically evaluate pp chain  and CN cycle rates  was undertaken in
1997-8, culminating in Solar Fusion I \citet[hereafter SFI]{sfi}.  Since SFI a
great deal of additional experimental and theoretical work has been done, e.g.
$^{14}{\rm  N(p,\gamma)^{15}O}$ \citep{luna:n14,imbriani,runkle,  bemm, lemut,
  marta},  $^{3}{\rm  He(^4He,\gamma)^{7}Be}$  \citep{singh,luna:s34,  gyurky,
  confortola, brown, erna:s34}, $^3\mathrm{He}(^3\mathrm{He},2\mathrm{p})^4$He
\citep{bonetti},   and   $^{7}{\rm   Be(p,\gamma)^{8}B}$   \citep{hammache:98,
  hammache:01, strieder:01, jung:02, jung:03,baby:03a, baby:03b, jung:10}. The
progress includes new facilities for  measuring cross sections at low energies
in or near  the Gamow peak, concerted efforts by  independent teams to address
key  uncertainties  such  as  the  S-factor  for  $^{7}\mathrm{Be}(\mathrm{p},
\gamma)^8$B, and  the development of  {\it ab initio} methods  for theoretical
extrapolations to solar  energies.  In order to assess the  impact of this new
work  on the  solar  model and  other  applications, a  second evaluation  was
recently undertaken, establishing new recommended values and uncertainties for
pp  chain  and CN  cycle  cross  sections  \citep[hereafter Solar  Fusion  II;
  SFII]{sfii}.   Here  we   present  new  solar  models  in   which  the  SFII
recommendations are adopted as new standard values, thereby providing the most
current  best  values  for  SSM  neutrino flux  predictions  and  other  solar
properties.

Standard  solar models  (SSMs) are  calibrated against  a  present-day surface
$(Z/X)_\odot$ value: a standard assumption  in SSMs is the equivalence between
solar   core   primordial    metal   abundances   and   contemporary   surface
abundances\footnote{The equivalence  is, of  course, {\it modulo}  the changes
  produced  by  microscopic diffusion.}.   Early  work  on SSMs  incorporating
reduced  CNO abundances \citep{bsp:2004,basu:2004,montalban:2004}  showed that
such solar models  are typically problematic: the location  of the boundary of
the convective envelope ($R_{\rm CZ}$),  the sound speed and density profiles,
and the present-day  surface helium abundance ($Y_{\rm S}$)  predicted by SSMs
using   low  $(Z/X)_\odot$  all   differ  from   the  values   extracted  from
helioseismology      by     amounts     beyond      accepted     uncertainties
\citep{montecarlo:2006}. On  the other  hand, SSMs using  higher $(Z/X)_\odot$
values,  e.g.  from  GS98,  show  excellent agreement  with  all  helioseismic
constraints  \citep{modelS, schlattl:1999,  bp01, bs:2005}.   The disagreement
between  `low-Z'   solar  model  predictions   and  helioseismic  constraints,
reminiscent of the now solved  `solar neutrino problem', is popularly known as
the `solar abundance problem'.

Various attempts  have been  made to solve  the solar abundance  problem.  For
example,    solar    abundance    determinations    have    been    questioned
\citep{antia:2005,antia:2006,  bbs:2005,delahaye:2006, delahaye:2010}  and the
constitutive     physics    of     solar    models     has     been    revised
\citep{basu:2004,montalban:2004,  bahcall:2005,guzik:2010}.  If we  assume the
correctness of the low-Z metallicity  determination for the Sun, there is not,
at  the moment,  a  satisfactory solution  that  brings the  solar model  into
agreement with helioseismology.  The only  way to achieve agreement in a low-Z
model  comparable to  that achieved  with  high metallicity  (GS98) models  is
through  {\it  ad  hoc} adjustments,  such  as  an  increase in  opacities  to
compensate  for the  lower metal  abundance  in the  solar radiative  interior
\citep{jcd:2009}.

An   alternative    possibility   was   considered    by   \citet{guzik:2006},
\citet{castro:2007}, and  more recently by  \citet{guzik:2010}.  These authors
assumed  that, soon  after the  Sun settled  on the  Main Sequence  (MS), very
metal-poor (or -free) matter was accreted onto the convective envelope, mixing
with and  diluting its primordial  composition. Solar models  were constructed
with    low-metallicity   AGSS09-like    convective   envelopes,    but   with
high-metallicity  radiative interiors  similar to  GS98.  It  was hypothesized
that  such  a   model  might  have  an  internal   structure  consistent  with
helioseismology,  despite  its  low-metallicity photosphere.   However,  large
discrepancies  in model sound  speed profiles  and convective  envelope depths
persisted.   Additional modifications  to the  models  (extended overshooting,
turbulent mixing)  were explored, but the  difficulties remained, particularly
the discrepant sound speed profile below the convective envelope (see Figure 1
in \citealt{castro:2007}).

While  the possibility  of  solving  the solar  abundance  problem through  an
inhomogeneous  model  is intriguing  --  we  know  metal inhomogeneities  were
established late in the evolution  of the proto-planetary solar disk because a
large excess of metal was sequestered in the planets \citep{haxton:2008} -- it
may be difficult  to guess {\it a priori} the specific  structure such a model
should take.   The work reported  here is a  first attempt to examine  a broad
class of possible nonhomogeneous solar  models that could be generated through
accretion,  which we  model  dynamically.   We explore  a  range of  accretion
histories, varying  the mass and composition of  the gas as well  as the epoch
during  which  the  accretion  occurs.   We study  the  detailed  helioseismic
properties of these models as  well as their neutrino flux characteristics, to
determine  whether any  of the  models might  be compatible  with  the various
observational constraints.   We also consider  the implications of  the models
for  future proposed  neutrino tests  of core  composition, such  as  might be
possible through  detection of CN-cycle neutrinos  \citep{haxton:2008} -- what
classes  of accretion  models could  be directly  ruled out  by  such neutrino
measurements?  This paper establishes  a baseline for future accretion studies
that could be  more directly tied to recent  evidence for peculiar differences
in the  surface abundances  of the  Sun and similar  stars (solar  twins), and
possible  connections   these  differences  may  have  to   the  chemistry  of
proto-planetary  disks  \citep{twins1}.   These  additional  studies  will  be
presented elsewhere.

The  paper is  laid out  as follows.  In \S~\ref{sec:ssm}  we  present updated
calculations  of  SSMs  for  high  and  low  metallicity  solar  compositions,
incorporating the  most recent nuclear reaction rates.  These models represent
the  baselines against  which models  with accretion  are later  compared.  In
\S~\ref{sec:calc}  we  describe our  calculational  procedures: the  numerical
algorithm and the parameter space the calculations span. Results for accreting
solar  models   are  presented   in  \S~\ref{sec:results}  and   discussed  in
\S~\ref{sec:discussion}.   Final  remarks  and  conclusions are  presented  in
\S~\ref{sec:conclusions}.

%
%

\section{Standard Solar Models} \label{sec:ssm}

The   new   SSM  calculations   presented   here   were   done  with   GARSTEC
\citep{garstec:2008},  including the  updates and  modifications in  the input
physics described in  \citet{ssm:09}.  The new models, in  addition, adopt the
nuclear reaction rates that were recently recommended in SFII.

SFII represents the first  systematic evaluation of hydrogen burning reactions
since the  SFI \citep{sfi} and  NACRE \citep{nacre} efforts.  SFII  provides a
set of standard S-factors and  uncertainties that reflect the progress made in
laboratory  and theoretical nuclear  astrophysics over  the last  decade.  The
team  of evaluators  carefully considered  which data  sets  were sufficiently
complete and  well documented to  be included; how  to combine data  sets that
span different energy ranges and  that might be somewhat inconsistent in their
error-bar  claims; and how  to associate  reasonable uncertainties  to fitting
functions (when the data sets span the region of astrophysical interest) or to
theoretical  extrapolations of data  (when no  data exist  in the  solar Gamow
peak).  The treatment of uncertainties was conservative, due to the increasing
relevance of the SSM as a  laboratory for weak interaction studies and stellar
evolution theory. The principal changes between SFII and SFI are:

\renewcommand{\labelitemi}{$\diamond$}
\begin{list}{\labelitemi}{\leftmargin=0.4cm}

\item  SFI  identified  $^7$Be(p,$\gamma)^8$B  as the  least  well  understood
  reaction  of   the  pp  chain,  and  called   for  additional  measurements,
  particularly below 300  keV.  The SFI evaluation was based  on a single data
  set \citep{filippone:a,filippone:b}  due to concern that  other results were
  either  too poorly  documented or  otherwise not  sufficiently  reliable for
  inclusion.  The  SFI recommended S-factor is  S$_{17}(0)=19^{+4}_{-2}$ eV b.
  The   value   used  in   previous   SSM   calculations  \citep{ssm:09}   was
  S$_{17}(0)=20.6$ eV  b \citep{jung:03}.  The SFII  evaluation benefited from
  four            new,            high-quality            data            sets
  \citep{hammache:98,hammache:01,strieder:01,jung:02,jung:03,baby:03a,baby:03b}
  and  yielded  S$_{17}$(0)=20.8  $\pm$  0.7  (expt)  $\pm$  1.4(theor).   The
  uncertainty  is dominated  by the  required theoretical  extrapolation.  The
  estimated total uncertainty at solar Gamow peak energies is 7.5\%.

\item  The   S-factor  for   the  controlling  reaction   of  the   CN  cycle,
  $^{14}$N(p,$\gamma)^{15}$O, was assigned  a large and asymmetric uncertainty
  in  SF  I,  S$_{114}$(0) =  3.5$^{+0.4}_{-1.6}$  keV  b.   A series  of  new
  measurements       ranging       from        70       to       480       keV
  \citep{luna:n14,imbriani,runkle,bemm,lemut,marta}  has led to  a significant
  reduction in the best value  and uncertainty: the SFII result, after summing
  all transitions,  is S$_{114}$(0)  = 1.66  $\pm$ 0.12 keV  b. This  value is
  close to  S$_{114}$(0) = 1.57 keV  b \citet{marta}, adopted  in our previous
  SSM  results \citet{ssm:09}.   The value  and uncertainty  are  important to
  future  plans to  use CN  neutrinos as  a probe  of the  solar-core C  and N
  abundances, as well as to age estimates for the galaxy's oldest stars.

 \item  Improved measurements  were  made  of the  reactions  that govern  the
   ppI/ppII+ppIII     branching     ratio,     $^3$He($^3$He,pp)$^4$He     and
   $^3$He($^4$He,$\gamma)^7$Be.   The SFI  uncertainties  for these  reactions
   were 7.4\%  and 9.4\%, respectively.  At  the time of SFI,  the most recent
   measurement of the $^3$He($^4$He,$\gamma)^7$Be  cross section was ten years
   old.   The new data  employed in  SFII for  $^3$He($^3$He,pp)$^4$He include
   measurements  that  extend  to the  lower  edge  of  the solar  Gamow  peak
   \citep{bonetti},  thus  removing  any   dependence  on  theory  apart  from
   screeening  corrections.   The  new  data  for  $^3$He($^4$He,$\gamma)^7$Be
   include       a       series       of       high-quality       measurements
   \citep{singh,luna:s34,gyurky,confortola,brown,erna:s34}  at  both high  and
   low energies,  spanning the range from  93 keV to 2.51  MeV.  The resulting
   SFII recommended S-factors are consistent with SFI values but significantly
   more precise,  S$_{33}$(0) = 5.21 $\pm$  0.27 MeV b and  S$_{34}$(0) = 0.56
   $\pm$   0.03   keV   b.    The    central   value   of   the   SFII   ratio
   S$_{33}(0)$/S$_{34}$(0)  is  8.7\%  lower  than  that  of  SFI,  increasing
   somewhat  ppII+ppIII  burning. Previously  \citep{ssm:09},  we had  adopted
   S$_{34}$(0) = 0.567  keV b \citet{constantini} and S$_{33}$(0)  = 5.4 MeV b
   (SFI).

 \item  The  S-factor for  the  driving reaction  of  the  pp chain,  p(p,e$^+
   \nu_e)$d, is  taken from theory,  as we have  no direct means  of measuring
   this reaction currently.  The SFI  value, S$_{11}(0) = 4.00 \cdot 10^{-22}$
   (1$\pm  0.007^{+0.020}_{-0.011})$ keV  b, includes  estimated uncertainties
   from input experimental quantities  such as the axial-vector coupling $g_A$
   and   theoretical   uncertainties   reflecting   possible   variations   in
   potential-model   calculations,  including  corrections   from  short-range
   physics and two-body currents.   The most significant development since SFI
   is the recognition that the overall nuclear physics uncertainty in S$_{11}$
   is governed  by a single low-energy  parameter which can  be constrained by
   using  the  precisely  know  tritium $\beta$  decay  rate  \citep{carlson}.
   Consequently  three  distinct  classes  of calculations,  potential  theory
   \citep{schiavilla}, hybrid  effective field theory  (EFT) \citep{park2003},
   and  pionless EFT  \citep{kong,butler,ando},  have now  converged in  their
   predictions.  The SFII recommended  S-factor is S$_{11}(0) = (4.01\pm 0.04)
   \cdot 10^{-22}$ keV  b, and the uncertainty in the  Gamow peak is estimated
   to be  0.9\%. The central value  is almost 2\% higher  than what originally
   \citet{park2003} recommended and was used in previous SSM calculations.

 \item The reaction $^3$He(p,e$^+ \nu_e)^4$He  is the source of the pp chain's
   most   energetic    neutrinos.    The   SFI   S-factor    best   value   is
   S$_\mathrm{hep}(0)  \sim 2.3  \cdot  10^{-20}$ keV  b,  but the  evaluators
   concluded  that no meaningful  uncertainty could  be assigned.   They noted
   that the scatter of theoretical  estimates corresponded to a factor of 2.5,
   up or  down.  After  SFI there were  significant efforts in  both potential
   models  \citep{marcucci} and  hybrid  EFT \citep{park2003}  to improve  the
   calculations  and  to assess  possible  sources  of  uncertainty.  The  two
   approaches  are   in  good  agreement.   The  recommended   SFII  value  is
   S$_\mathrm{hep} = (8.6 \pm 2.6) \cdot 10^{-20}$ keV b.

 \item New  data from  the LUNA collaboration  on $^{15}$N(p,$\gamma$)$^{15}$O
   \citep{bem09} that  agree with an  older data set from  \citet{hebbard} but
   not with  \citet{rr74} led  to the SFII  conclusion that  some unidentified
   systematic error may have affected  the low-energy data of \citet{rr74}.  A
   significantly  lower  value for  S$_{115}^\gamma(0)=36  \pm  6$  keV b  was
   recommended  as  an interim  value,  pending  the  outcome of  very  recent
   experiments now being analyzed.

 \item  The  SFII  ratio  of  pep  to  pp  rates  is  2.6\%  higher  than  the
   corresponding  SFI  ratio because  the  former  includes  a calculation  of
   radiative corrections \citep{kurylov} while the later was evaluated at tree
   level.
 \end{list}
 These new  rates, and others from  SFII not explicitly  discussed above, were
 incorporated in the new solar model calculations.

All  models   in  this  work   are  required  to  reproduce   the  present-day
($\tsun=4.57$~Gyr)        solar        luminosity,       $L_\odot=3.8418\times
10^{33}$~\mbox{erg~s$^{-1}$},       and      radius      $R_\odot=6.9598\times
10^{10}~\mbox{cm}$.  In  this section we present  two SSMs that  differ in the
present-day surface metal-to-hydrogen abundance ratios.  One model is based on
the GS98  solar composition, $\left(Z/X\right)_{\rm S}= 0.0229$,  the other on
AGSS09, $\left(Z/X\right)_{\rm S}= 0.0178$; in both cases the adopted scale is
meteoritic\footnote{Except  for  volatile  elements;  see  \citet{agss09}  for
  details.}.

In Table~\ref{tab:ssm}  we list the  most relevant parameters for  these SSMs.
The  models are  labeled according  to  the adopted  solar compositions.   The
results differ  little from those of  earlier models with  the same respective
compositions, presented in \citet[Table~2]{ssm:09}. The SFII reaction rates do
not introduce  noticeable changes  in the initial  composition of  the models,
denoted  here  by  the  initial  metal, $\zini$,  and  helium,  $\yini$,  mass
fractions.  Consequently,  the surface and  central mass fractions  of metals,
$\zs$ and $Z_{\rm  C}$, and of helium, $\ys$ and $Y_{\rm  C}$, are also almost
unchanged from those  of \citet{ssm:09}. The depth of  the convective envelope
$\rcz$ is  slightly deeper in the models  with the SFII rates  and the average
fractional rms deviation between the solar and the model sound speed profiles,
$\left< \delta  c/c\right>$, is also slightly  better in the  new models.  The
changes are, however, small compared to model uncertainties, so the new models
are effectively fully consistent with previous ones.

\begin{deluxetable}{lcc}
\tablewidth{0pt}
\tablecaption{Main characteristics of standard solar models.\label{tab:ssm}}
\tablehead{ & GS98 & AGSS09}
\startdata
$(Z/X)_{\rm S}$ & 0.0229 & 0.0178 \\
$Z_{\rm S}$ & 0.0170 & 0.0134 \\
$Y_{\rm S}$ & 0.2429 & 0.2319 \\
$R_{\rm CZ}/R_\odot$ & 0.7124 & 0.7231 \\
$\left< \delta c / c\right>$ & 0.0009 & 0.0037 \\
$Z_{\rm c}$ & 0.0200 & 0.0159 \\ 
$Y_{\rm c}$ & 0.6333 & 0.6222 \\ 
$Z_{\rm ini}$ & 0.0187 & 0.0149 \\
$Y_{\rm ini}$ & 0.2724 & 0.2620 \\
\enddata
\end{deluxetable}

Given the similarities between the new models and those of \citet{ssm:09}, the
reader is  referred to that  source for a  more detailed discussion.  Here, we
focus on manifestations of the solar abundance problem, the primary motivation
for  the  studies presented  here.   There are  three  key  indicators of  the
abundance problem,  the depth of  the convective envelope, the  surface helium
abundance, and the sound speed profile. The AGSS09 model produces a convective
zone that is too  shallow and a surface helium abundance that  is too low (see
Table~\ref{tab:ssm}).   The entries  in the  table should  be compared  to the
helioseismic values, $\rcz= 0.713\pm  0.001$ \citep{basu:1997} and $0.2485 \pm
0.0035$  \citep {basu:2004}, respectively.  If helioseismic  uncertainties are
combined  with   SSM  uncertainties  (\citealp   {montecarlo:2006},  equations
32~and~35),  then  AGSS09  model  predictions  differ  from  solar  values  at
$3\sigma$ for both  $\rcz$ and $\ys$. In contrast,  GS98 model predictions are
in  excellent agreement  with the  parameters extracted  from helioseismology.
Although more  difficult to  quantify, the difference  between the  AGSS09 and
solar sound  speed profiles, determined  from $\left< \delta c/c  \right>$, is
about four  times larger than that  obtained with the  higher metallicity GS98
model.   These specific  discrepancies for  the AGSS09  abundance set  are the
quantitative manifestations  of the solar  abundance problem mentioned  in the
introduction.

\begin{deluxetable}{lccc}
\tablewidth{0pt}
\tablecaption{Solar neutrino fluxes.\label{tab:neutrinos}} 
\tablehead{ $\nu$ flux & GS98 & AGSS09 & Solar\tablenotemark{a}} 
\startdata 
pp & $5.98(1 \pm 0.006)$ & $6.03(1 \pm 0.006)$ & $6.05(1^{+0.003}_{-0.011})$ \\
pep & $1.44(1 \pm 0.012)$ & $1.47(1 \pm 0.012)$ & $1.46(1^{+0.010}_{-0.014})$\\
hep & $8.04(1 \pm 0.30)$ & $8.31(1 \pm 0.30)$ & $18(1^{+0.4}_{-0.5})$\\
$^7$Be & $5.00(1 \pm 0.07)$ & $4.56(1 \pm 0.07)$ & $4.82(1^{+0.05}_{-0.04})$\\
$^8$B & $5.58(1 \pm 0.14)$ & $4.59(1 \pm 0.14)$ & $5.00(1\pm 0.03)$\\ 
$^{13}$N & $2.96(1 \pm 0.14)$ & $2.17(1 \pm 0.14)$ &$\leq 6.7$\\ 
$^{15}$O & $2.23(1 \pm 0.15)$ & $1.56(1 \pm 0.15) $ &$\leq 3.2$\\ 
$^{17}$F & $5.52(1 \pm 0.17)$ & $3.40(1 \pm 0.16)$ & $\leq 59.$\\
$\chi^2/P^{\rm agr}$& 3.5/90\% & 3.4/90\%& --- \\
\enddata
\tablecomments{Neutrino   fluxes   are   given  in   units   of
    $10^{10}$(pp), 
$10^9$($^7$Be), $10^8$(pep, $^{13}$N, $^{15}$O), $10^6$($^8$B, $^{17}$F) 
and $10^3$(hep)~\mbox{cm$^{-2}$ s$^{-1}$}.  Asymmetric  uncertainties have been
averaged.}
\tablenotetext{a}{Solar neutrino  fluxes inferred from  all available neutrino
  data \\ \protect\citep{borex:2011}.}
\end{deluxetable}

The new SFII nuclear reaction rates do alter the
predicted solar neutrino fluxes, particularly the fluxes associated
with the ppIII chain and CN-cycle, mechanisms for solar hydrogen burning that
are relatively unimportant energetically.
 With  respect to  the rates  used  in \citet{ssm:09},  the  most
important SFII changes are
the  2\% increase  in ${\rm  S_{11}(0)}$, the 3\% decrease in ${\rm S_{33}(0)}$, 
and the 5\%  increase in ${\rm S_{1\, 14}(0)}$.  Other changes affecting
predicted neutrino fluxes include the reduced proton capture cross section on
$^{15}$N and the increase in the pep/ pp rate
ratio. Model predictions for  the neutrino fluxes and associated uncertainties
are presented in
Table~\ref{tab:neutrinos} for both the AGSS09 and GS98 compositions.  
The most significant changes  are  a  5\% decrease  in  the
predicted $^8$B  flux primarily because of the increase in ${\rm
  S_{11}}$ and  the increase in  the $^{13}$N flux  due to the  larger central
abundance  of C.  The increase in C is a consequence of the lower SFII value
for $^{15}$N(p,$\gamma)^{16}$O, a reaction that competes with the CN I cycle
reaction $^{15}$N(p,$\alpha)^{12}$C and allows mass to flow out of the
CN I cycle into CN II.  
We discuss the parametric dependence of the fluxes on nuclear cross sections
and other SSM input parameters elsewhere, including some of the resulting
neutrino tests that might be made of solar composition and S-factors \citep{PGSH}. 
 
 Table~\ref{tab:neutrinos}  also includes  the updated  solar  neutrino fluxes
 inferred from all available neutrino  data. The analysis includes the recent
 more precise,  $^7$Be measurement, which is  the main change  with respect to
 previous analysis \citep{roadmap,bps08,concha:2010}.  Details of the analysis
 will  be shown  elsewhere \citep{borex:2011}.  In  order to  compare the  SSM
 predictions with the fluxes inferred from neutrino data, we use the $\chi^2$ 
 function   defined   in   \citet{concha:2010},   with  updated   errors   and
 correlations. 
We find $\chi^2_{\rm GS98}=3.5$ and $\chi^2_{\rm AGSS09}=3.4$, leading in both
cases to $P^{\rm agr}_{\rm GS98,AGSS09}=90\%$. The new fusion
cross sections  from SFII  and the  new Borexino results  lead to  both models
predicting  solar  neutrino  fluxes   in  excellent  agreement  with  inferred
ones;  from Table~\ref{tab:neutrinos}  it can  be seen  that solar  $^7$Be and
$^8$B  fluxes are  intermediate between  SSM predictions  for GS98  and AGSS09
compositions.  Currently,  solar   neutrinos  can  not  differentiate  between
different solar compositions.

%
%

\section{Solar Models With Accretion: Computational Details} \label{sec:calc}

\subsection{The algorithm}\label{sec:algo}

Briefly,  the construction  of an  SSM  requires finding  two parameters  that
determine  the   initial  composition  of  the  model,   $\yini$  and  $\zini$
($\xini=1-\yini-\zini$), and  the mixing length  parameter $\alpha_{\rm MLT}$.
The  three parameters  are determined  iteratively by  requiring the  model to
reproduce   the    present-day   solar   luminosity,    radius   and   surface
metal-to-hydrogen abundance ratio.  We remind the reader that  the choice of a
solar composition  determines not only  the present-day $\left(Z/X\right)_{\rm
  S}$  value, but  also  the relative  abundances  of metals  within the  bulk
metallicity. In what  follows, all models are computed  using the AGSS09 solar
composition.

The  algorithm  used to  construct  SSMs can  be  easily  modified to  compute
nonstandard models with accretion, once  the properties of the accreted matter
(mass and composition)  and the timescales and accretion  rates are specified.
Below, and in  the remainder of the paper, we  use the following nomenclature:
$\mac$ is the  mass of the accreted material, $\zac$,  $\xac$, and $\yac$ are,
respectively, its metal, hydrogen and helium mass fractions, $\tau_{\rm ac,i}$
and  $\tau_{\rm ac,e}$  are  the initial  and  ending times  of the  accretion
process.  Schematically, the algorithm consists of the following steps:

\begin{enumerate}

\item An initial pre-main sequence (PMS) model of mass $M= \msun - M_{\rm ac}$
  is constructed using initial  guesses for $\yini$, $\zini$, and $\alpha_{\rm
    MLT}$.

\item The  model is  evolved at constant  mass until $\tau=  \tau_{\rm ac,i}$.
  The accretion  then begins  at a constant  rate $\dot{M}$,  continuing until
  $\tau=\tau_{\rm ac,e}$, when $M= \msun$. $\zac$ is a parameter of the model,
  while  $\xac$ and  $\yac$  are  determined by  the  conditions $\xac/\yac  =
  \xini/\yini$  and $1=\xac  +  \yac +  \zac$.  That is,  we  assume that  the
  hydrogen-to-helium ratio  of the accreted  material is identical to  that of
  the initial solar material.

\item From $\tau=\tau_{\rm ac,e}$ onward the model is evolved at constant mass
  until $\tau=\tsun$.  Model properties  are checked against present-day solar
  values,  and the differences  are used  to improve  the initial  guesses for
  $\yini$, $\zini$, and $\alpha_{\rm MLT}$.  The process is repeated until the
  solar surface properties are reproduced to the desired accuracy (typically 1
  part in $10^4$ or better).

\end{enumerate}

In the above scheme the free parameters  of the model are the same as those of
SSMs, namely $\yini$, $\zini$, and  $\alpha_{\rm MLT}$. But they now depend on
properties of the accreted material, particularly its composition, because the
mixture  of initial  and  accreted  matter that  forms  the present-day  solar
surface must satisfy the $\left(Z/X\right)_{\rm S}$ constraint.

So  far, we  have assumed  that  the parameters  characterizing the  accretion
process  are known.   We now  discuss  how we  selected the  ranges for  those
parameters.

\subsection{Models}\label{sec:grids}

The  properties of  the  resulting solar  models  will clearly  depend on  the
parameters that  describe both  the properties of  the accreted mater  and the
timing  and duration  of  the accretion  process.   Our goal  was  to cover  a
relatively  large parameter  space, in  order to  better define  the  range of
possible solar model outcomes.  Below we describe our choices.

\subsubsection{Metallicity of the accreted mass}

Intuitively, a hypothetical solution to the solar abundance problem would seem
to require  that metal-poor or even  metal-free material is  accreted onto the
Sun.   This  has  been  the  assumption  in  the  models  explored  previously
\citep{castro:2007,guzik:2010}.  Here  we explore metal-free,  metal-poor, and
metal-rich accretion.  Surprisingly,  as described in \S~\ref{sec:results}, we
find  that  metal-rich accretion  can  lead  to  partial improvements  in  the
helioseismic properties of some models.  We explore $\zac$ values ranging from
0 to 0.03, the latter being  roughly twice the solar surface metallicity.  The
relative abundances  of metals is not  varied, but is fixed  to AGSS09 values.
This is  an assumption  that may  be a reasonable  starting point  for initial
explorations, but probably should be relaxed in future calculations given that
any  mechanism  for  segregating  metals  would  likely  reflect  condensation
temperatures and other element-specific chemistry.

\subsubsection{Accreted mass}

We consider accretion of mass  from the protoplanetary disk with a composition
different from  that of  the unprocessed material  of the  proto-solar nebula.
The  amount  of  such material  the  Sun  might  have  accreted in  its  early
evolutionary phases in not well constrained,  as estimates of both the mass of
the proto-planetary  disk and the fraction  of that mass that  remained in the
solar system  vary.  Assuming  that initially the  proto-solar cloud  was well
mixed, one can set  a lower limit to the mass of  the protoplanetary disk, the
so-called minimum-mass solar nebula (MMSN)  limit, from estimates of the metal
content  of the  planets.  The  calculation, however,  still depends  on model
assumptions.   Early estimates  \citep{weiden:1977,hayashi:1981}  give $M_{\rm
  MMSN} \approx 0.013  - 0.07~{\rm M}_\odot$, assuming planets  were formed at
their  present locations.  More  sophisticated models  that take  into account
migration give somewhat larger  values. \citet{desch:2007} finds $M_{\rm MMSN}
\approx  0.17~{\rm  M}_\odot$   but  with  a  disk  that   is  being  strongly
photo-evaporated, i.e.,  most of its  mass migrates outwards, not  towards the
Sun.

Based on these  considerations, we explore values of $\mac$  in the range from
$3\cdot  10^{-5}$  to  0.06~$\msun$,  or  equivalently,  from  10  to  $2\cdot
10^4\,\mearth$.  We stress that this is the total accreted mass, not the metal
mass.  The upper  end of the range thus  comes within a factor of  2-3 of most
estimates of $M_  {\rm MMSN}$.  Given that models of  massive disks with steep
density   profiles  lead   to   substantial  \citep{desch:2007}   evaporation,
0.06~$\msun$ is probably  a safe upper bound on the amount  of mass that could
have been chemically  processed in the protoplanetary disk  and later accreted
onto the  Sun.  In terms of the  total accreted metal, our  choices for $\mac$
and $\zac$ yield a range of 0 to 600~$\mearth$.

\subsubsection{Accretion timescales and mass accretion rate}\label{sec:time}

\begin{figure}[h]
\includegraphics[scale=.55]{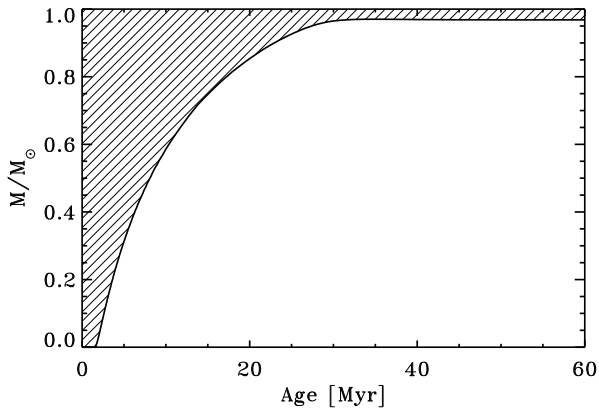}
\caption{Time dependence of $M_{\rm CE}$ during early evolutionary phases of a
  solar  model;   the  shaded  area  depicts  the   convective  zone.  $M/{\rm
    M_\odot}=1$ defines the surface of the model.  \label{fig:mconv}}
\end{figure}

The  time when accretion  takes place  is of  particular importance.   This is
clear from  inspection of Figure~\ref{fig:mconv},  where the evolution  of the
mass of the  convective envelope, $M_{\rm CE}$, during  early SSM evolution is
shown. Initially,  the model is fully  convective on the Hayashi  track. As it
evolves towards the  Zero Age Main Sequence (ZAMS),  a radiative core develops
and convection  quickly retreats.  During these early  phases, the $e$-folding
time for the  evolution of $M_{\rm CE}$ is about 8~Myr.   At about 30~Myr, the
star  reaches  the  ZAMS  and  further  evolutionary  changes,  including  the
extension of the convective envelope, occur on very long (nuclear) timescales.
For a fixed  amount of accreted material, the mass  of the convective envelope
at the  time of accretion determines  the degree of  dilution (or enrichment).
These  considerations  lead  us   to  identify  three  paradigmatic  accretion
scenarios: \renewcommand{\labelitemi}{$\diamond$}
\begin{list}{\labelitemi}{\leftmargin=0.4cm}

\item  {\em early accretion}:  Accretion starts  when the  convective envelope
  still  encompasses  most of  the  star;  accreted  material is  then  highly
  diluted.  We take  $\tau_{\rm ac,i}=5$~Myr as a representative  time for the
  early  accretion  scenario.  In  typical  models,  the  convective  envelope
  contains 70 to 80\% of the total solar mass at this time.

\item  {\em  intermediate accretion}:  Accretion  starts  when the  convective
  envelope has  receded towards the surface  and contains about 20  to 30\% of
  the stellar mass. Accreted material  is strongly diluted in the envelope but
  the effects of  accretion in the final composition of  the model are clearly
  noticeable.   We take  $\tau_{\rm  ac,i}=15$~Myr as  representative of  this
  case;

\item {\em late  accretion}: Accretion starts when the model  is very close to
  the ZAMS, i.e.  when timescales for structural changes begin to be dominated
  by nuclear burning  timescales. The convective envelope is  mature, close to
  its  final  MS  size,  and  changes  very little  afterwards.  Its  mass  is
  comparable to  some of the  largest $\mac$ values we  consider. Accordingly,
  the final  structure of models with  late accretion is  strongly affected by
  the properties of the accreted material. We take $\tau_{\rm ac,i}=30$~Myr as
  representative of late accretion.

\end{list}

Accretion could  occur at even later  times. However, given  that $M_{\rm CE}$
changes very  little after the first  30~Myr of evolution,  such models should
not  differ significantly  from the  {\em late  accretion}  scenario described
above.  In terms of involved  timescales, the {\em late accretion} scenario is
equivalent to that investigated by \citet{castro:2007} and \citet{guzik:2010},
who assumed accretion took place once the Sun had settled on the MS.

Computationally, the accretion  scenarios are defined in terms  of the time at
which accretion takes place.  Physically, however, the relevant characteristic
is the thickness  of the convective envelope. Therefore,  as long as accretion
occurs sufficiently early that element diffusion effects are still negligible,
the  thickness of  the convective  envelope  is the  only physically  relevant
parameter.  For this reason, even if  the young Sun avoided a fully convective
Hayashi phase \citep{wuch:2001,baraffe:2010} and consequently developed a thin
convective   envelope  after   only  a   few  Myr   \citep{baraffe:2010},  our
intermediate  and  late accretion  results  remain  valid, constraining  these
alternative evolutionary tracks.

The duration of the accretion phase ($\dtau$) is a second important parameter,
particularly for the early and  intermediate cases, where $M_{\rm CE}$ evolves
rapidly.  Our standard choice is $\Delta \tau_{\rm ac}=10$~Myr.  There are two
reasons  for this  choice.   One  is simplicity:  this  choice corresponds  to
quiescent   accretion  rates   well   below  $10^{-7}~\msun$~\mbox{yr$^{-1}$},
allowing us to  neglect the energy deposition on the outer  layers of the star
and associated effects that could  alter final internal structure of the model
\citep{baraffe:2009}.  We assume that  the central  star is  almost completely
assembled when  accretion from the  protoplanetary disk, that is  accretion of
chemically   processed   material,  begins.    The   second   reason  is   the
correspondence to  typical estimates of the planetary  formation time, usually
in the  range of a few  to 10 Myr.  However,  to test the  impact of different
accretion  durations,  we have  computed  models  with  more rapid  accretion,
$\dtau=1$~Myr.

In all cases considered, accretion is assumed to occur at a constant mass rate
$\dot{M}= \mac/\dtau$.

\begin{figure}
\includegraphics[scale=.55]{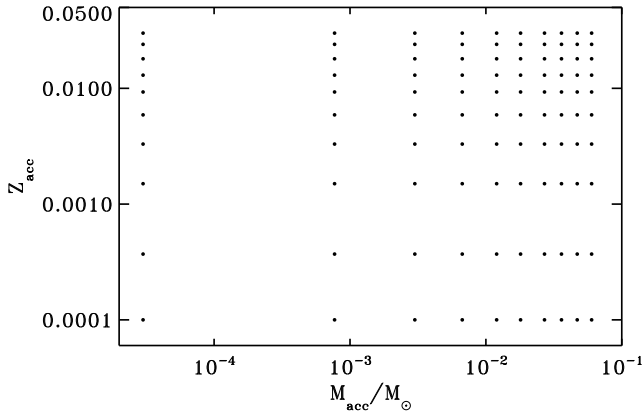}
\caption{The mass and metallicity grid for the model calculations.
\label{fig:grid}
}
\end{figure}

\subsection{Grid of models}

The ranges considered for $\mac$  and $\zac$ were explored by stepping through
a 10 $\times$ 10 grid, with  quadratic spacings.  Therefore, for each value of
$\tau_{\rm ac,i}$ and $\Delta \tau_{\rm  ac}$, 100 models were generated. This
basic grid is depicted in Figure~\ref{fig:grid}.

%
%

\section{Solar Models With Accretion: Results}
\label{sec:results}

Here we  present results  for the three  accretion scenarios  discussed above,
early, intermediate, and late.  We first focus on the relationship between the
altered solar composition and helioseismic observables, then later discuss the
effects on neutrino fluxes.

\subsection{Early accretion}

\begin{figure*}[ht]
\includegraphics[scale=.67]{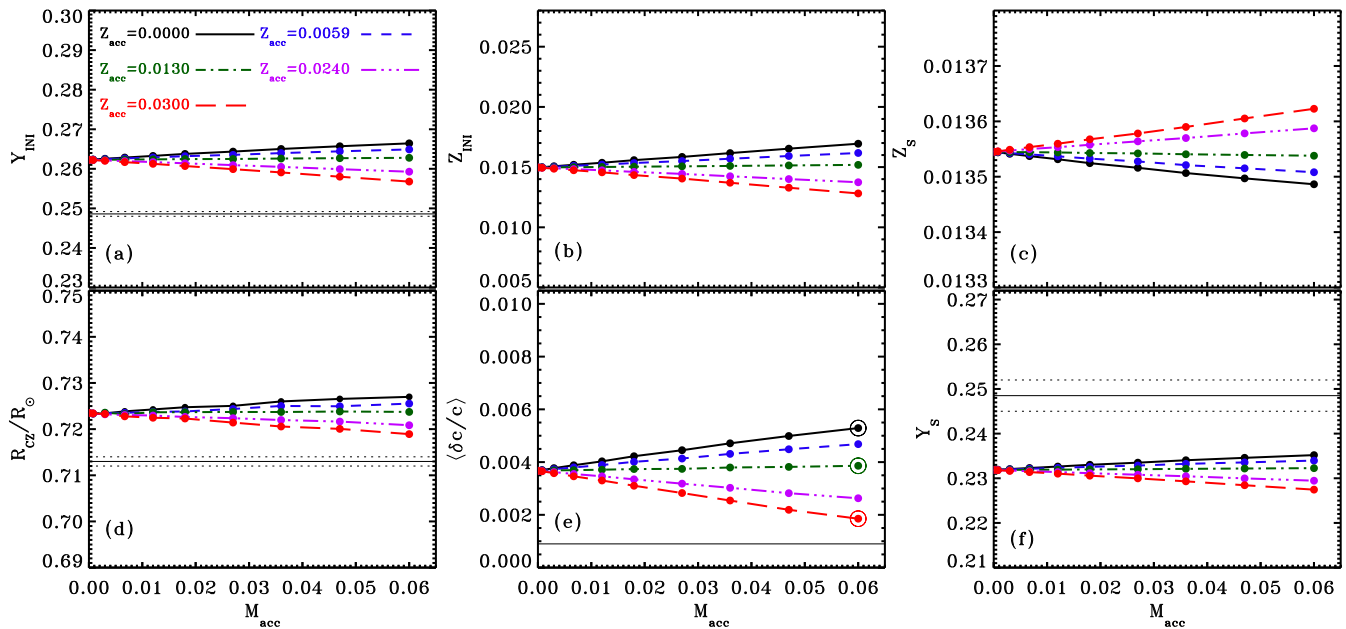}
\caption{Composition and helioseismic quantities of sequences of accreting 
solar models with $\tau_{\rm   ac,i}=5$~Myr   and   $\Delta   \tau_{\rm
  ac}=10$~Myr (early accretion).  Results  are shown for five different values
of $\zac$, as labeled in panel $a$. 
\label{fig:hv0a5t10}
}
\end{figure*}

\begin{figure}[t]
\includegraphics[scale=.55]{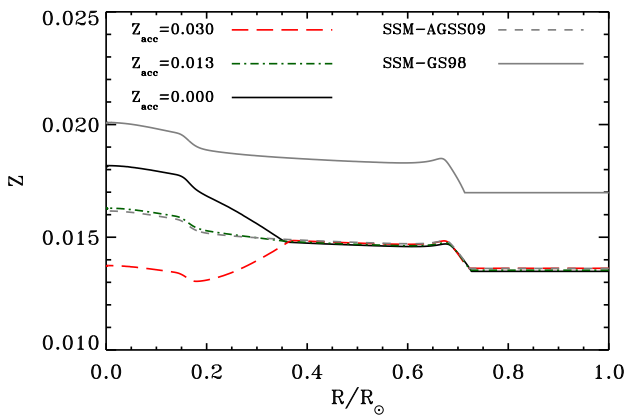}
\caption{Final  metallicity  profile  as  a  function  of  radius,  for  three
  accreting models with the $\zac$ values indicated in the plot and for
  $\mac=0.06$~$\msun$. For  comparison, the SSM metallicity  profiles are also
  shown (see \S~\ref{sec:ssm}).  
\label{fig:zprofa05}
}
\end{figure}

\begin{figure}[t]
\includegraphics[scale=.55]{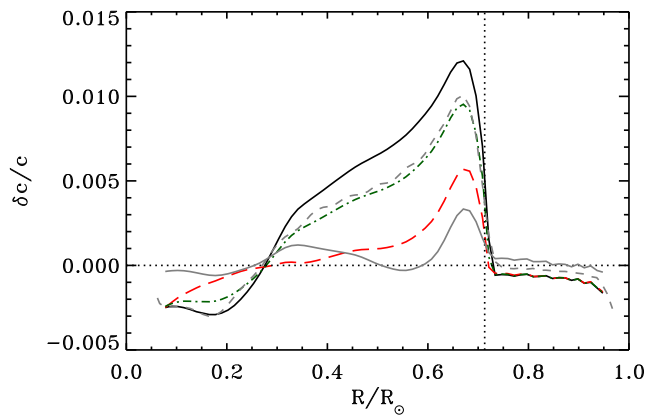}
\caption{Sound speed profiles for the models of Figure~\ref{fig:zprofa05}. The
  vertical dotted line marks the boundary of the convective envelope.  
\label{fig:dcv00a05t10} 
}
\end{figure}

\subsubsection{Composition and helioseismic quantities}

Model  compositions  and  resulting   helioseismic  quantities  are  shown  in
Figure~\ref{fig:hv0a5t10}  for  five  selected  model  sequences  of  constant
$\zac$, from metal-free  to super-solar metallicity.  Panels {\em  a} and {\em
  b}  give  the  initial  composition  of  the  models,  $\yini$  and  $\zini$
respectively, i.e.   the composition before accretion takes  place. Panel {\em
  c} shows  the present-day surface  metallicity $\zs$. The lower  panels show
selected solar  properties that can  be directly tested  with helioseismology,
$\rcz$ in panel {\em d}, $\dc$ in panel {\em e}, and $\ys$ in panel {\em f}.

We first examine the behavior of  the initial composition of the models. Since
all models have to satisfy $(Z/X)_{\rm S}=0.0178$, metal-poor accretion models
lead to higher $\zini$ values.   A higher metallicity core implies an increase
in  the   radiative  opacity  and  molecular  weight,   i.e.,  larger  central
temperatures.  However,
models are constrained by $\lsun$ as  well. So, an increase in nuclear burning
caused by  higher central temperatures has  to be compensated  by limiting the
amount of  fuel available,  that is  increasing $\yini$. In  panel {\em  a} we
include  the   primordial  helium  mass   fraction  from  Standard   Big  Bang
Nucleosynthesis (SBBN),  $Y_{\rm P}=0.2486 \pm 0.0006$, using  the latest WMAP
results for the baryon-to-photon rato $\eta$ \citep{wmap:2010}.

The   effects   of   accretion   on   metallicity  profiles   are   shown   in
Figure~\ref{fig:zprofa05} for  three models with $\mac=0.06$.   The models are
identified in panel {\em e} by  the open circles.  For comparison, we included
the results for  the SSMs presented in \S~\ref{sec:ssm}.   The strong gradient
in the metallicity  profile between $\sim 0.17$ and  0.35~$\rsun$ seen for the
metal-rich and metal-free models is  the result of mixing between the accreted
and the primordial material, while the convective zone is receding towards the
surface.  As discussed  above, the surface metallicity is  very similar in all
models and nearly equal to that of the AGSS09 SSM.

Model values for  $\rcz$, $\dc$, and $\ys$, quantities  tightly constrained by
helioseismology,  are  shown in  panels  {\em  d}, {\em  e},  and  {\em f}  of
Figure~\ref{fig:hv0a5t10},  respectively.  Panels  {\em  d} and  {\em f}  also
include the  helioseismic values for  $\rcz$ and $\ys$, plotted  as horizontal
solid lines,  and their  $1\sigma$ uncertainties, shown  as dotted  lines.  We
first  discuss the  behavior of  $\rcz$ as  a function  of $\zac$.   The outer
layers  (above 0.8~$\msun$, see  Figure~\ref{fig:zprofa05}) of  models evolved
with metal-poor accretion have lower metallicities.  Additionally, lower $\zs$
requires  lower $\xs$. Both  effects tend  to lower  the radiative  opacity at
$\rcz$  \citep{bsp:2004},  producing  a  shallower convective  envelope.   The
opposite is true  for metal-rich accretion. Therefore, the  behavior of $\rcz$
seems  to argue  against metal-free  (or metal-poor)  accretion.  In contrast,
metal-rich  accretion leads  to  deeper convective  envelopes,  closer to  the
helioseismic value.   The trends in the  behavior of the  sound speed profiles
are quite similar, as one sees from  panel {\em e}.  There the GS98 SSM result
is shown as a horizontal line. The  behavior of $\dc$ and $\rcz$ are very well
correlated  because the bulk  of the  discrepancy in  the sound  speed profile
comes from  the mismatch between  the solar and  model $\rcz$ values.  To show
this,  we  plot  in  Figure~\ref{fig:dcv00a05t10}  the  relative  sound  speed
difference  as   a  function   of  radius  for   the  same  models   shown  in
Figure~\ref{fig:zprofa05}  and   identified  in   panel  {\em  e}   with  open
circles. The bump  below the convective zone decreases  for models with $\rcz$
close to the  solar value (vertical dotted line).   This occurs for metal-rich
accretion  models.  The SSM  results  for  AGSS09  and GS98  compositions  are
included for comparison.

Unfortunately, the behavior of $\ys$  is qualitatively opposite that of $\rcz$
and $\dc$.  Metal-poor accretion leads to  higher $\ys$ values as  a result of
larger  $\yini$  \citep{solarhelium:2010}, improving  the  agreement with  the
helioseismic determination.  In contrast, metal-rich accretion leads to lower,
more  discrepant,  $\ys$ values.  Therefore,  helium abundance  considerations
favor metal-poor accretion.
 
\begin{figure*}[ht]
\includegraphics[scale=.67]{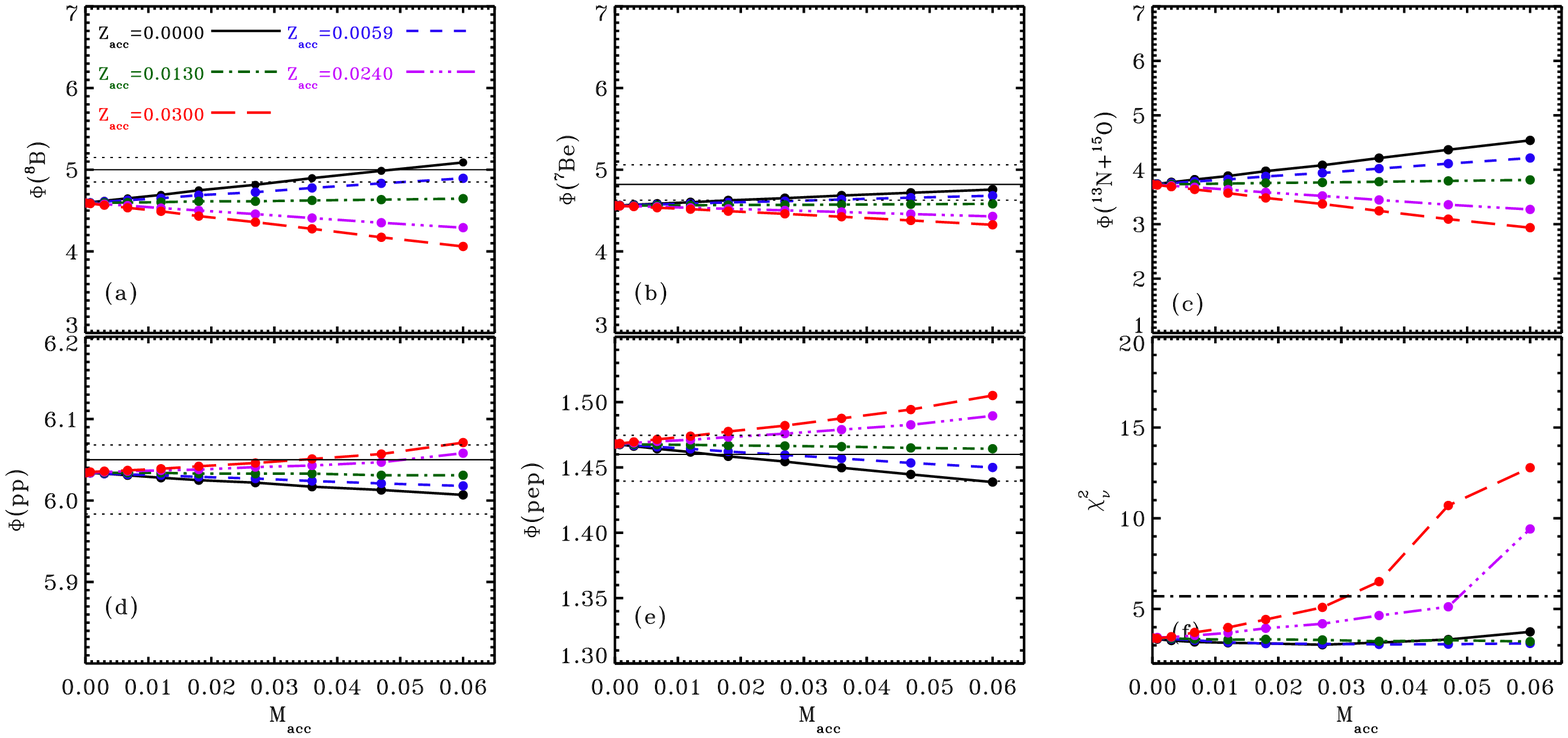}
\caption{Neutrino    fluxes    for    the    accreting   solar    models    of
  Figure~\ref{fig:hv0a5t10}. 
\label{fig:nuv0a5t10}
}
\end{figure*}

These results  suggest that it may  not be possible to  construct solar models
with early accretion in which  all predicted helioseismic variables agree with
solar values  extracted from  helioseismic observations.  While  agreement for
$\rcz$ and $\dc$  can be improved with metal-rich  accretion, the accompanying
lower  $\ys$  values increase  discrepancies  for  that variable.   Similarly,
metal-poor  accretion helps  with $\ys$,  but increases  the  discrepancies in
$\rcz$ and $\dc$.

\subsubsection{Neutrino fluxes}

Solar  neutrino fluxes place  a second  set of  constraints on  solar interior
properties.  We  will compare  model predictions to  the flux  values 
determined from a global analysis of solar neutrino experiments
\citep{borex:2011}  that are
given in the last column of Table~\ref{tab:neutrinos}.  The authors 
considered all  available data from solar neutrino  experiments, including new
determinations  of  the  $^8$B  flux  by  SNO  \citep{sno:2010}  and  Borexino
\citep{boreb8:2010} and the most recent measurement of the $^7$Be flux 
by Borexino \citep{borebe7:2008,borex:2011}.

The general dependence of solar  model fluxes on accretion simply reflects the
initial   metallicity,  which   determines  the   temperature  in   the  Sun's
energy-producing  core.   Fluxes  that  depend more  strongly  on  temperature
($^8$B, $^{13}$N, $^{15}$O, $^{17}$F and, to a lesser extent, $^7$Be) increase
with increasing core metallicity  (metal-poor accretion). However, because the
models are constrained by $L=L_\odot$,  neutrino fluxes less dependent on core
temperature (pp, pep and hep) must decrease
\citep{nutemp:1996}.  In our accretion  models the dependence of each neutrino
flux on solar core metallicity can be characterized by the partial derivatives
\begin{equation}
\frac{\partial \log{\Phi(\nu_i)}}{\partial \log{Z_{\rm ini}}}= \alpha(\nu_i).
\end{equation}
We determine $\alpha(\nu_i)$, for each  flux $i$, by performing a joint linear
fit to all early accretion models. The fluxes most strongly correlated
with core  temperature yield  $\alpha(^7{\rm Be})= 0.34$,  $\alpha(^8{\rm B})=
0.82$, $\alpha(^{13}{\rm  N}+^{15}{\rm O})= 1.58$.   For pp and pep  fluxes we
find  $\alpha(\mathrm{pp})= -0.036$,  $\alpha(\mathrm{pep})= -0.16$.   In what
follows,  the  sum of  the  fluxes  $^{13}{\rm  N}+{}^{15}\mathrm{O}$ will  be
denoted CN.

\begin{figure*}[ht]
\includegraphics[scale=.67]{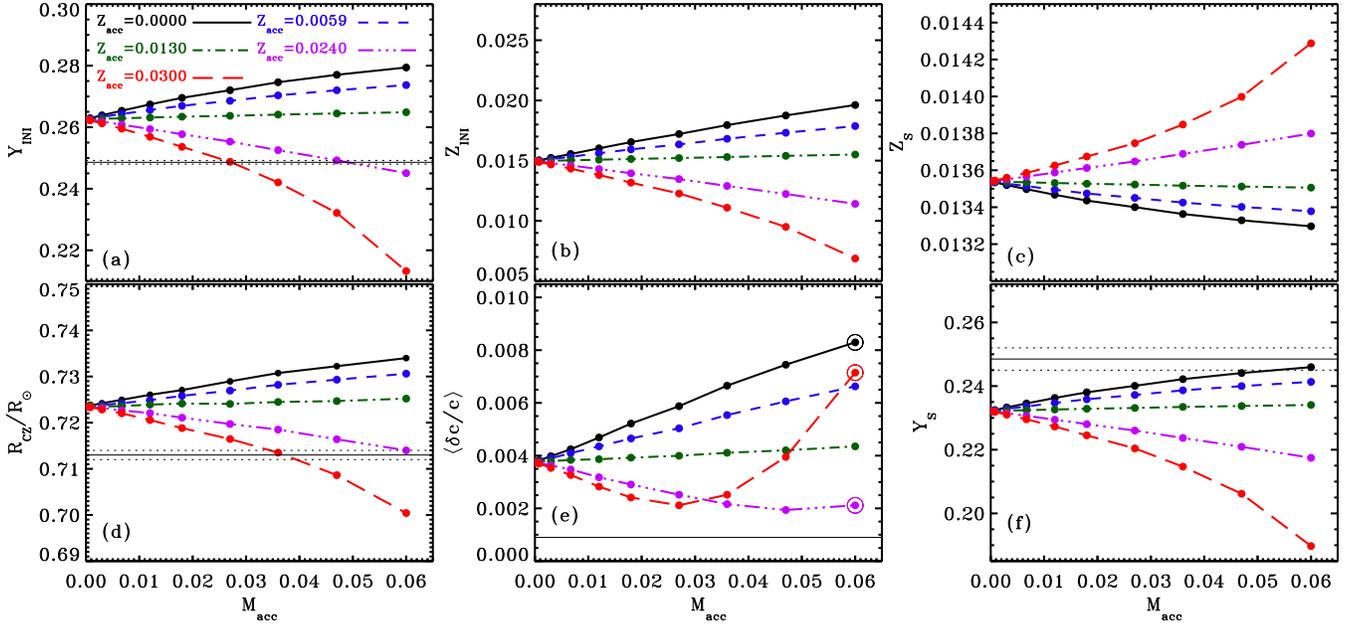}
\caption{Composition and helioseismic quantities of sequences of accreting 
solar   models   with    $\tau_{\rm   ac,i}=15$~Myr   and   $\Delta   \tau_{\rm
  ac}=10$~Myr (intermediate accretion).  Results  are shown for five different
values of $\zac$, as labeled in panel $a$. 
\label{fig:hv0a15t10}
}
\end{figure*}

Results    for     the    various    neutrino    fluxes     are    shown    in
Figure~\ref{fig:nuv0a5t10}.  The  metallicity sequences  are  the  same as  in
Figure~\ref{fig:hv0a5t10}.   Panels {\em  a} and  {\em b}  show the  $^8$B and
$^7$Be fluxes,  which have been  measured directly in experiments  (the $^7$Be
flux  can be  extracted  from  Borexino results  given  other measurements  of
neutrino   oscillation   parameters).   The   solar  fluxes   (third   column,
Table~\ref{tab:neutrinos})  are shown with  horizontal black  lines, including
$1\sigma$  uncertainties.  Models  with metal-rich  accretion tend  to deviate
from the solar values, particularly for  the precisely known $^8$B flux due to
its  sensitivity  to core  temperature.   As  mentioned  before, the  greatest
sensitivity to accretion is exhibited by  the combined CN flux (panel {\em c})
because, in addition to its  dependence on core temperature, this flux depends
linearly on  the total  C+N abundance in  the solar core.   Unfortunately, the
current  determination  of   the  combined  flux  is  only   an  upper  limit,
$\Phi(\mathrm{CN}) \leq  9.9 \times 10^8~{\rm  cm^{-2} s^{-1}}$, and  does not
allow to put any constraints on the solar core metallicity. 

Panels {\em  d} and {\em  e} show the  strongly correlated pp and  pep fluxes.
The luminosity  constraint and the  $^7$Be measurement by Borexino  combine to
produce  the  small  uncertainties  in  these fluxes  (about  1\%  and  1.5\%,
respectively).   As with  $\Phi(^8$B),  particularly the  pep flux somewhat
disfavors  metal-rich accretion  models,  especially those  with large  $\mac$
values. 

As an  indicator of the overall  agreement between model and  solar fluxes, we
calculated, for each model, the $\chi^2$ function (see \S~\ref{sec:ssm}).  The
results are shown in panel {\em f}.  The 
horizontal dashed-dotted line shows  the 68.3\% probability band.  As expected
from  results for  individual fluxes,  metal-poor accretion  models lead  to a
small overall  improvement in the  agreement with experimental  fluxes, mostly
driven  by improvements  in the  agreement with  $\Phi(^8$B)  and $\Phi$(pep).
Note that the sequence
with $\zac= 0.0130$  shows almost no changes in  the predicted neutrino fluxes
and $\chi^2$ values: this value of $\zac$ is
nearly equal  to the AGSS09 $\zini$  value, so that the  accreted material and
initial solar  material have nearly  identical compositions.  We  caution that
these neutrino  flux changes should not  be overinterpretted, as  we have seen
that models  with metal-poor accretion tend  to do poorly on  $\rcz$ and $\dc$
and therefore their  internal structure shows differences with  respect to the
Sun. 

\subsection{Intermediate accretion}

\begin{figure}[ht]
\includegraphics[scale=.55]{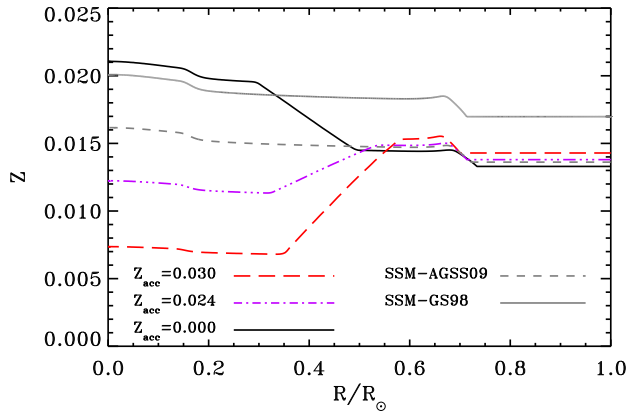}
\caption{Metallicity profiles for selected models (open circles in panel {\em
    e}  in Figure~\ref{fig:hv0a15t10}).   SSM  results are  also  shown, as  in
  Figure~\ref{fig:zprofa05}.  
\label{fig:zprofa15}
}
\end{figure}

\subsubsection{Composition and helioseismic quantities}

The results  for $\yini$, $\zini$, and $\zs$  are shown in panels  {\em a, b},
and  {\em  c}, respectively,  of  Figure~\ref{fig:hv0a15t10}. The  qualitative
trends track those we observed in early accretion models: metal-poor accretion
leads to higher  $\zini$ and $\yini$ and, because of  the lower $\xini$, lower
$\zs$.   Metal-rich  accretion  produces  the opposite  behavior.   Metal-rich
accretion for $\zac=0.024$  and 0.030 and with accreted  masses above 0.05 and
0.027~$\msun$, respectively,  lead to $\yini$  lower than the  SBBN primordial
value $Y_{\rm P}$.   The same models lead to $\zini$ values  that are low when
compared to metallicities observed in the solar neighborhood (see
e.g.   \citealt{przy:2008}).   Compared   to  early   accretion   models,  the
consequences of accretion are amplified in the intermediate models because the 
convective envelopes are shallower at the time of accretion.
The  large  contrast  between  $\zini$ and  $\zs$  for
intermediate  accretion  models  is apparent  from  Figure~\ref{fig:zprofa15},
where the metallicity profiles for the three models identified with open circles
in  Figure~\ref{fig:hv0a15t10} are shown.   The  partially mixed  region
extends  roughly  between  0.35  and  0.55~$\rsun$, depending  on  the  model.
Profiles of  the GS98  and AGSS09 SSMs  are also shown.   Metal-rich accretion
models,  as stated above,  lead to  under-metallic cores.   The consequences for
resulting model sound-speed profiles are significant, as discussed
below.

\begin{figure}
\includegraphics[scale=.55]{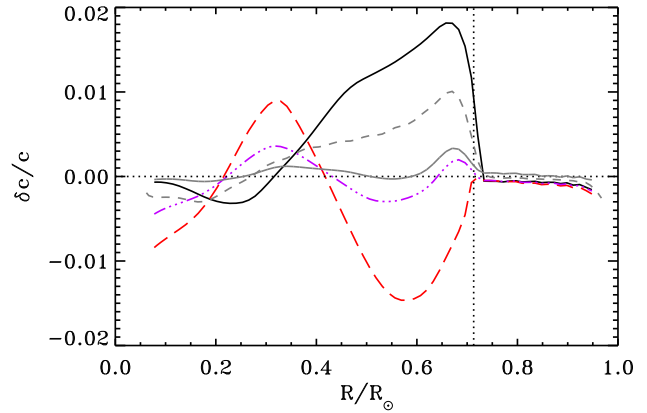}
\caption{Sound  speed  profiles  for the models  of
  Figure~\ref{fig:zprofa15}. 
\label{fig:dcv00a15t10}
}
\end{figure}

\begin{figure*}[ht]
\includegraphics[scale=.67]{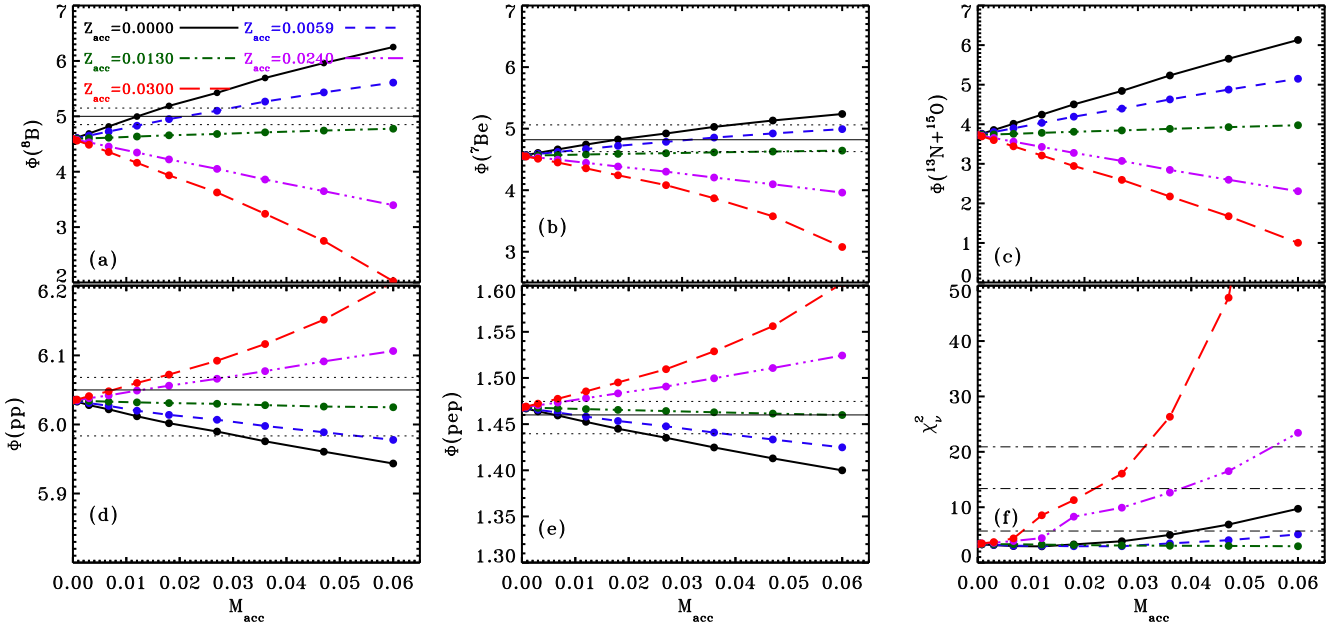}
\caption{Neutrino    fluxes    for    the    accreting   solar    models    of
  Figure~\ref{fig:hv0a15t10}. 
\label{fig:nuv0a15t10}
}
\end{figure*}

Results for the helioseismic quantities  $\rcz$, $\dc$, and $\ys$ are shown in
panels {\em  d, e}, and  {\em f} of  Figure~\ref{fig:hv0a15t10}, respectively.
The qualitative behavior of $\rcz$ and $\ys$ is quite similar to that found in
the early  accretion models.  Interestingly, models with  metal-rich accretion
can lead to $\rcz$ in excellent agreement with the helioseismic value. However,
the  same models  yield $\ys$  values that  are low  compared  to helioseismic
values, as well as $\yini$ values below the SBBN value.  For $\zac=0.030$, the
highest  $\mac$ values considered,  the convective  envelopes are  deeper than
allowed by helioseismology.  This is reflected in the  behavior of $\dc$.  For
the $\zac=0.03$ sequence, $\dc$  has a minimum around $\mac=0.027~\msun$.  The
agreement is better than with the AGSS09  SSM, but still a factor of two worse
than with the GS98 SSM.   The agreement deteriorates for larger $\mac$ values,
as the convective envelope becomes too deep. 
Figure~\ref{fig:dcv00a15t10}  gives  the sound  speed  profiles  for the  same
models that  were used in Figure~\ref{fig:zprofa15}.  The  sound speed profile
for the  model with metal-free  accretion (black solid line)  is significantly
degraded compared  to that of AGSS09  SSM, because the  convective envelope is
too shallow;  the bump  below the  convective zone is  very large  and extends
quite deep, to $\sim 0.35~\rsun$. The difficulty exists despite the similarity
between  the  metallicity  of  this  model  and that  of  AGSS09  SSM  over  a
substantial  region,   0.45-0.7~$\rsun$,  below  their   convective  envelopes
(Figure~\ref{fig:zprofa15}   for    $M>0.9\msun$).    For   $\zac=0.024$   and
$\mac=0.06\msun$  (green line-dotted  line),  the model  predicts the  correct
$\rcz$, so  that the bump  below the convective  envelope is almost  gone.  At
small $r$, the sound speed  difference oscillates around zero, and the overall
$\dc$ is twice as  large as for the GS98 SSM (shown  as the grey dashed line).
Finally, the model with $\zac=0.030$, which we noted had too deep a convective
zone, exhibits a large bump in  the sound speed profile of opposite sign below
the convective zone, as well as significant discrepancies at smaller radii due
to its under-metallic interior. 

\subsubsection{Neutrino fluxes}

Relative  to early  accretion models,  the neutrino  fluxes  from intermediate
accretion models exhibit slightly  more sensitivity to $\zini$: $\alpha(^7{\rm
  Be})=   0.52$,  $\alpha(^8{\rm   B})=  1.12$,   $\alpha({\rm   CN})=  1.77$,
$\alpha(\mathrm{pp})=   -0.045$,  and  $\alpha(\mathrm{pep})=   -0.14$.   Flux
results are  shown in  Figure~\ref{fig:nuv0a15t10}. The higher  sensitivity of
temperature-dependent fluxes, particularly  $\Phi(^8$B), $\Phi(^7$Be), and the
combined  $\Phi(\mathrm{CN})$,  reflects the  decreased  dilution of  accreted
material.  Panel  {\em a}  shows that $\Phi(^8$B),  in models  with metal-rich
accretion,  differs markedly  from the  solar value  (horizontal  black line).
Conversely,  metal-poor   accretion  modestly  improves   the  agreement  with
experiment  for almost  all the  range  of $\mac$  we considered.   Metal-free
accretion can 
also  lead to an  improved $\Phi(^8$B),  but only  for smaller  $\mac$ values.
Otherwise, for models with $\mac \gtrsim 0.04\msun$, the predicted $\Phi(^8$B)
is  too large.   Because $^7$Be  (panel  {\em b})  is less  sensitive to  core
temperature  changes  (and less  precisely  known), significant  discrepancies
arise only  for metal-rich  accretion at large  $\mac$ values.  Panel  {\em c}
shows the sensitivity of the combined  CN flux to the accretion, but this flux
is too poorly constrained experimentally to allow any conclusions to be drawn.

The results  for the pp and  pep fluxes (panels  {\em d} and {\em  e}), though
dependent  on  the luminosity  constraint,  strongly  suggest that  metal-rich
accretion can be excluded  for $\mac \gtrsim 0.03~\msun$.  However, metal-free
and metal-poor accretion, over the  range of $\mac$ values we considered, lead
to models  that are  perfectly consistent  with the pp  and pep  solar values.
This conclusion is  consistent with the one we drew  from examining the $^7$Be
flux.

The global  $\chi^2$ in panel  {\em f} shows  that, as in the  early accretion
case, the  metal-free and metal-poor accretion scenarios  can slightly improve
the overall 
agreement between the  models and the the solar neutrino  fluxes. It should be
recalled once  more, however,  that both AGSS09  and GS98 SSMs  give excellent
agreement with experimentally determined sloar neutrino fluxes
(Table~\ref{tab:neutrinos}).  The horizontal lines in the plot are the 68.4\%,
5\%,  and 0.1\%  probability levels.   Comparing  the $\chi^2$  values of  the
different $\zac$  sequences with results  for the individual  neutrino fluxes,
one  finds  that  the  $^8$B  flux,  with its  small  uncertainty  and  strong
temperature  sensitivity, dominates  the $\chi^2$.  For example,  although the
model with  $\zac=0.00$ and  $\mac=0.06\msun$ yields nearly  perfect agreement
with the  $^7$Be, pp, and pep  fluxes, a relatively high  $\chi^2$ is obtained
because the model overestimates the $^8$B neutrino flux.

\subsection{Late accretion}\label{sec:late}

\begin{figure*}[t]
\includegraphics[scale=.67]{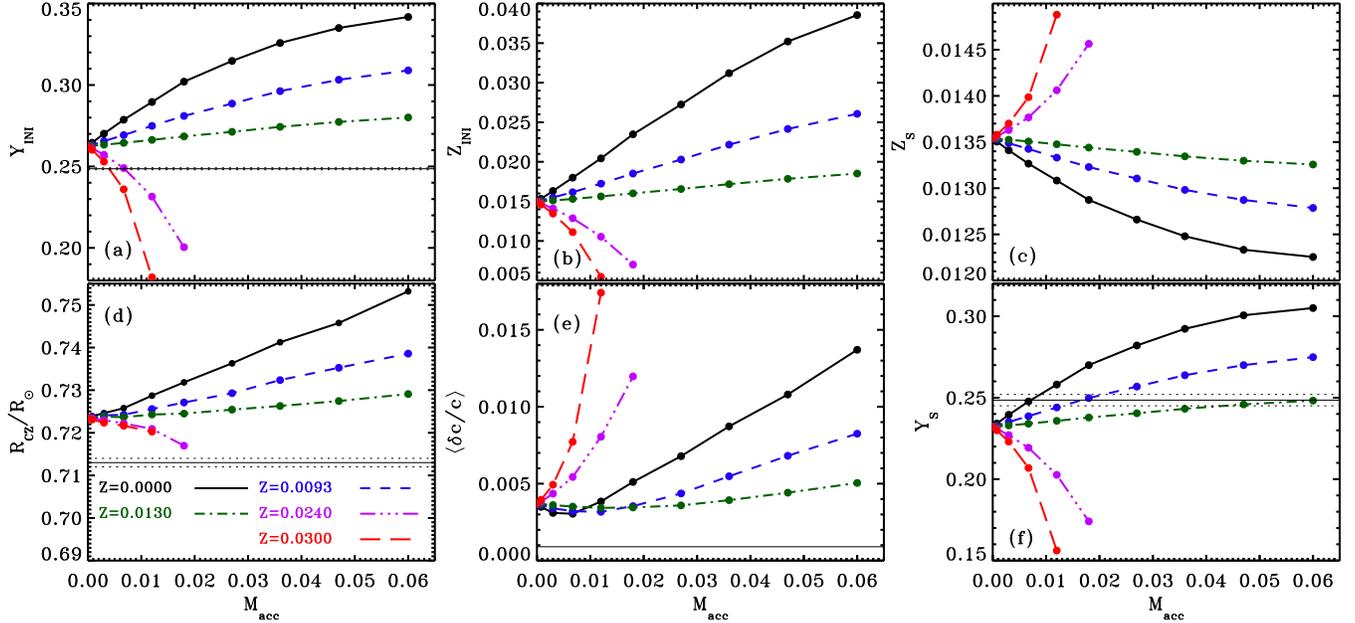}
\caption{Helioseismological quantities  and initial composition  for accreting
  solar models with $\tau_{\rm ac,i}=30$~Myr and $\Delta \tau_{\rm ac}=10$~Myr
  (late accretion).   Results are shown  for five different values  of $\zac$,
  labeled in panel $d$.  
\label{fig:hv0a30t10}
}
\end{figure*}

\begin{figure*}[t]
\includegraphics[scale=.67]{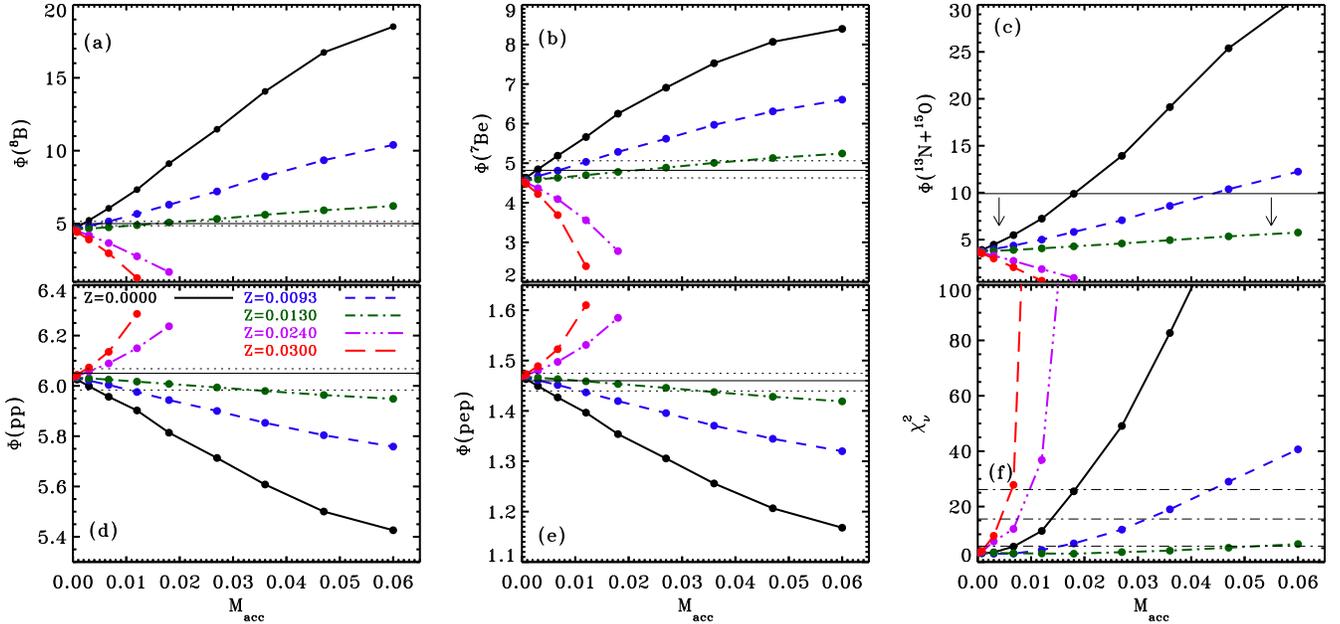}
\caption{Neutrino   fluxes  for  the accreting   solar  models  of
  Figure~\ref{fig:hv0a30t10}.
\label{fig:nuv0a30t10}
}
\end{figure*}

\subsubsection{Composition and helioseismic quantities}

In a late accretion model the accretion occurs when the Sun is close to the MS
with a well-developed radiative core and consequently a
thin  convective envelope.   The effects  of the  accretion on  the convective
envelope are therefore maximal.

The composition and helioseismic properties of the models are summarized in
Figure~\ref{fig:hv0a30t10}.    The  sequences   for  $\zac=0.024$   and  0.030
terminate at  $\mac=0.02$ and 0.012$\msun$  respectively.  As is  evident from
the plot of $\zini$ in panel {\em b}, larger values of $\mac$ are incompatible
with modern Sun's $(Z/X)_{\rm S}$: the needed initial metallicity 
is driven to {\em negative} values.  Even for smaller, allowed
$\mac$ values,  metal-rich accretion  leads to very  low $\yini$  values, well
below the  SBBN value, as shown  in panel {\em a}.   

The helioseismic  quantities $\rcz$,  $\dc$, and $\ys$  are plotted  in panels
{\em  d,  e},  and  {\em  f}, respectively.   The  metal-free  and  metal-poor
accretion  models (solid  black and  blue  short dashed  lines) yield  shallow
convective envelopes.  Also, for  relatively low $\mac$ values ($\mac \lesssim
0.015~\msun$), the overall  sound speed profiles, as estimated  from $\dc$ and
$\ys$, are somewhat improved. These  models have initial compositions and thus
core compositions very similar to that of the GS98 SSM. However, as the models
do poorly on $\rcz$, discrepancies in the sound speed profiles remain.  Models
with larger $\mac$ values at low  $\zac$ fail to reproduce the observed $\rcz$
and  $\dc$, and  also  overestimate $\ys$.   We  discuss metal-poor  accretion
further in \S~\ref{sec:cmpothers}, where  we compare our results with previous
work. 

In contrast to early  and intermediate  accretion models, sound-speed profiles
for  late,  metal-rich  models  do  not  improve,  regardless  of  the  $\mac$
value.  The  accretion's  dramatic  effects  on surface  composition  lead  to
unacceptably  low values  $\zini$ values  even for  small $\mac$  values.  The
resulting  low-metallicity cores  are not  compatible with  the radiative-zone
sound speed profiles.  $\ys$ is also too low for these models.

\subsubsection{Neutrino fluxes}

Neutrino fluxes are quite sensitive  to parameter variations in late accretion
models, responding to the large changes in $\zini$. The power-law exponents we
derive   are:   $\alpha(^7{\rm  Be})=   0.66$,   $\alpha(^8{\rm  B})=   1.43$,
$\alpha({\rm     CN})=    2.08$,     $\alpha(\mathrm{pp})=     -0.081$,    and
$\alpha(\mathrm{pep})= -0.18$.

The individual fluxes are given in Figure~\ref{fig:nuv0a30t10}.  The agreement
between  model  and  observed  fluxes  deteriorates  in  both  metal-free  and
metal-rich accretion models.   This is mostly driven by  the $^8$B flux (panel
{\em a}),  but the other  fluxes with relatively small  uncertainties ($^7$Be,
pp, and  pep) also  tend to  trend away from  observation.  Thus  the neutrino
fluxes limit the  accreted composition to near-SSM values,  independent of the
$\mac$  value.   The  slightly   metal-poor  $\zac=  0.0093$  sequence  is  an
interesting one, yielding for $\mac \sim 0.01\msun$ a somewhat improved fit to
the  observed  fluxes.   In this  model  $\dc$  and  $\ys$ are  also  somewhat
improved, while the  convective envelope is only slightly  shallower than that
of the AGSS09 SSM.

\section{Discussion} 
\label{sec:discussion}

We  have investigated whether  the solar  abundance problem  can be  solved or
alleviated by  constructing solar models with different  interior and envelope
compositions.   Rather than postulating  an {\it  ad hoc}  modern Sun  with an
inhomogeneous structure, we instead generate that structure through a physical
mechanism,   accretion  of   metal-poor  or   metal-rich  material   from  the
protoplanetary  disk.    This  mechanism   is  well  motivated   both  because
significant accretion is  seen in T Tauri stars, which  may resemble the early
Sun,  and because  we  see  clear evidence  in  our solar  system  of the  the
large-scale segregation  of metal  that accompanied planetary  formation.  The
chemical processes that concentrated dust  and ice at the disk's midplane, and
the physical process of planetary formation that scoured those metals from the
disk, locking them into the  planets, could have combined to create conditions
for  either metal-poor  creation accretion  (deposition of  metal-depleted gas
from the  disk's outer surface layers) or  metal-rich accretion (proto-planets
spiraling into  the Sun) after the  Sun had developed its  radiative core.  In
this way the composition of the Sun's convective zone could have been altered,
becoming chemically distinct from the Sun's radiative core, which would have a
composition quite close to that of the solar system's primordial gas.

By  exploring  accretion  as   the  candidate  mechanism  responsible  for  an
inhomogeneous modern Sun, we can  generate nonstandard models as a function of
physical parameters, such as the mass  and composition of the accreted gas and
the  time of  accretion.  The  accretion  parameters, and  thus the  resulting
models,  are constrained  to reproduce  the  composition of  the modern  solar
surface, taken to be  AGSS09 in our models.  We can then  test these models by
comparing  their  properties  against  contemporary  solar  observations  done
through either helioseismology or solar neutrino spectroscopy.  The goal is to
determine whether  one can account for  modern observations that  appear to be
contradictory in  the context  of the  SSM -- the  solar abundance  problem --
through  a plausible physical  mechanism that  could have  negated one  of the
SSM's key assumptions, a homogeneous ZAMS Sun.

The possibility that  metal-poor accretion could be responsible  for the solar
abundance problem  has motivated earlier  work \citep{castro:2007, guzik:2006,
  haxton:2008,  nordlund:2009, guzik:2010}.   Our work  extends  these earlier
efforts  by  exploring a  broad  range  of  accretion histories,  masses,  and
compositions, and by testing the  resulting models for consistency with modern
observations  of  the  Sun's  surface  and interior.   However,  there  remain
unexplored degrees of freedom: we  have assumed a constant mass transfer rate,
a  fixed duration  of the  accretion, and  fixed composition  of  the accreted
material. We discuss our findings and their implications below.

\subsection{Helioseismic quantities}

The  most apparent  manifestations  of  the solar  abundance  problem are  the
inconsistencies between  observations and SSM solar  surface helium abundances
(too low), convective envelope depths (too shallow), and the solar sound speed
profiles,  particularly in the  upper radiative  zone.  The  main helioseismic
results for our updated SSMs can be found in Table~\ref{tab:ssm}.

How do the  model predictions change if we  postulate metal-rich or metal-poor
accretion?  Let us first consider results  for $\rcz$ and $\ys$ for the early,
intermediate, and late  accretion scenarios, shown in panels  {\em d} and {\em
  f}      of     Figures~\ref{fig:hv0a5t10},      \ref{fig:hv0a15t10},     and
\ref{fig:hv0a30t10}, respectively.   Qualitatively, metal-poor accretion leads
to higher $\ys$ values and  shallower convective envelopes with respect to the
SSM. The  opposite happens for metal-rich accretion.  Therefore, accretion can
be invoked  to construct models that  match the helioseismic  values of either
$\ys$ (metal-poor) or $\rcz$ (metal-rich), but at the expense of enlarging the
discrepancy in the other quantity. No model simultaneously improves both $\ys$
and $\rcz$.

\begin{figure}[t]
\includegraphics[scale=.7]{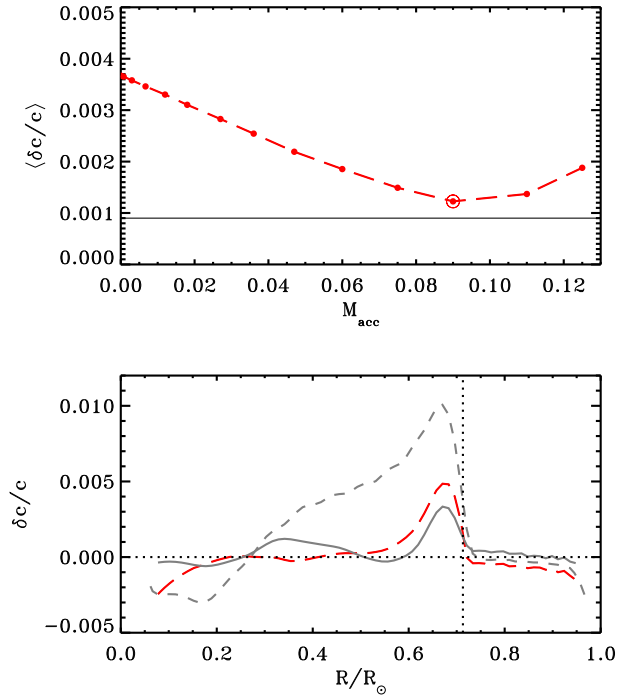}
\caption{Top panel: average rms sound speed for the extended sequence of early
  accretion models,  for $\zac=0.030$. Best model,  for $\mac=0.090~\msun$, is
  marked with  an open circle. Bottom  panel: the red long-dashed  line is the
  sound speed profile  for the $\mac=0.090~\msun$ model. The  GS98 (grey solid
  line)  and  AGSS09  (grey   short-dashed)  SSM  profiles  are  included  for
  comparison.  
\label{fig:z30ext} 
}
\end{figure}

The discrepancies we observe between  solar and model sound speed profiles are
correlated with the discrepancies in $\rcz$.  Better agreement in $\rcz$ leads
in general  to smaller  $\dc$ values.   This is a  consequence of  $\dc$ being
dominated by  the large  bump in  the sound speed  difference right  below the
convective  envelope  that is  accentuated  whenever  there  is a  significant
mismatch between the solar and model  values for $\rcz$.  The best results for
intermediate (late) accretion lead to  $\dc$ being about two (three) times the
value  for the  GS98 SSM.   Interestingly, the  best results  for intermediate
accretion    are   from    a    metal-rich   case    (panel    {\em   e}    in
Figures~\ref{fig:hv0a15t10}) and for late accretion are from a metal-poor case
(panel  {\em  e}  in  Figures~\ref{fig:hv0a30t10}).  The  latter  provides  an
exception to  the general rule of  correlated $\dc$ and  $\rcz$, stemming from
the sensitivity of  $\zini$ in late accretion models to  the properties of the
accreted material. We discuss this further in \S~\ref{sec:cmpothers}.

For the  early accretion scenario, the  results displayed in panel  {\em e} of
Figure~\ref{fig:hv0a5t10}  for  metal-rich   accretion  suggest  that  further
improvement in $\dc$ might be achieved  if $\mac > 0.06~\msun$.  To test this,
we  computed models  with $\zac=0.030$  and $\mac=$~0.075,  0.090,  0.110, and
0.125~$\msun$.   The results  for $\dc$  are given  in Figure~\ref{fig:z30ext}
(top panel).  A  minimum is found around $\mac=0.090~\msun$,  yielding a value
for $\dc$  close to that of the  GS98 SSM (solid black  horizontal line).  For
accretion masses greater than  $\mac=0.090~\msun$, $\dc$ begins to rise again.
In the bottom  panel the sound speed profile  for the $\mac=0.090~\msun$ model
is shown (red long-dashed line).  It is  similar to that of the GS98 SSM (grey
solid  line), except  close to  the Sun's  center, where  the  metallicity and
helium abundances in the model are too low. The AGSS09 SSM sound speed profile
is also shown.  For this  model $\rcz= 0.716\, \rsun$, in reasonable agreement
with the  helioseismic value $0.713\pm0.001\, \rsun$.   However, following the
general relation between $\rcz$ and $\ys$ previously described, this model has
$\ys=0.225$, further from  the helioseismic value, $\ys=0.2485\pm0.0035$, than
the AGSS09  SSM value ($\ys=0.2319$,  the limit of no  accretion).  Therefore,
larger accretion masses do not offer  a global solution to the solar abundance
problem.

All  calculations presented  in Section~\ref{sec:results}  have  been computed
assuming  a  fixed duration  for  the  accretion  phase of  $\Delta  \tau_{\rm
  ac}=10$~Myr. Estimated lifetimes of protoplanetary disks are about a few Myr
\citep{haisch:2001,williams:2011} so that longer timescales for accretion seem
unlikely.  On  the other hand,  shorter timescales are possible,  perhaps even
favored.   We therefore  repeated the  calculations  from \S~\ref{sec:results}
using  $\Delta  \tau_{\rm ac}=1$~Myr,  to  test  the  consequences of  shorter
accretion  times.  As  one might  expect  intuitively from  an examination  of
Figure~\ref{fig:mconv},  the effect of  a reduced  $\Delta \tau_{\rm  ac}$ for
given  $\zac$, $\mac$,  and $\taui$  is  analogous to  enhancing the  dilution
factor  of  the accreted  material:  that  is,  for two  otherwise  equivalent
accretion scenarios  with the  same starting times,  the one with  the shorter
duration will deposit material into  a larger mean convection-zone mass during
the  accretion, and  thus that  material will  experience more  dilution.  For
example,  results  for ($\taui$,  $\Delta  \tau_{\rm ac}$)=(15~Myr,1~Myr)  are
intermediate between  those for (5~Myr,10~Myr ) and  (15~Myr,10~Myr), shown in
Figures~\ref{fig:hv0a5t10}  and \ref{fig:hv0a15t10},  respectively.   For late
accretion, e.g.,  after the model has settled  on the MS, the  duration of the
accretion phase is irrelevant, provided  it is short compared to gravitational
settling timescales:  at late times, there  is little evolution  in either the
depth or  the mass of the convective  zone.  Thus we conclude  that changes in
the  duration  of  the accretion  will  not  alter  our basic  conclusion:  no
combination of the accretion variables $\mac$, $\zac$ and $\taui$ will lead to
a model in which all helioseismic predictions are improved.

\subsection{Solar neutrinos}

We have  compared model solar neutrino  fluxes with solar fluxes  (column 3 in
Table~\ref{tab:neutrinos}).  The  latter have  been  derived  from a  combined
analysis  of all solar neutrino experiments with the addition of the
solar luminosity constraint\footnote{The luminosity constraint equates the
solar luminosity to the rate of energy generation in the core, assuming a
steady-state Sun.} \citep{borex:2011}.
The analysis includes  the CN-cycle fluxes,  although only upper limits
are currently available.   We find, using the new  nuclear reaction rates from
SFII and the  newest Borexino results for the  $^7$Be flux \citep{borex:2011},
that  the GS98  and AGSS09  SSMs are  both in  excellent agreement  with solar
neutrino data, producing comparable fits. These
results are summarized in the last row of Table~\ref{tab:neutrinos}.

\begin{figure*}[th]
\includegraphics[scale=.67]{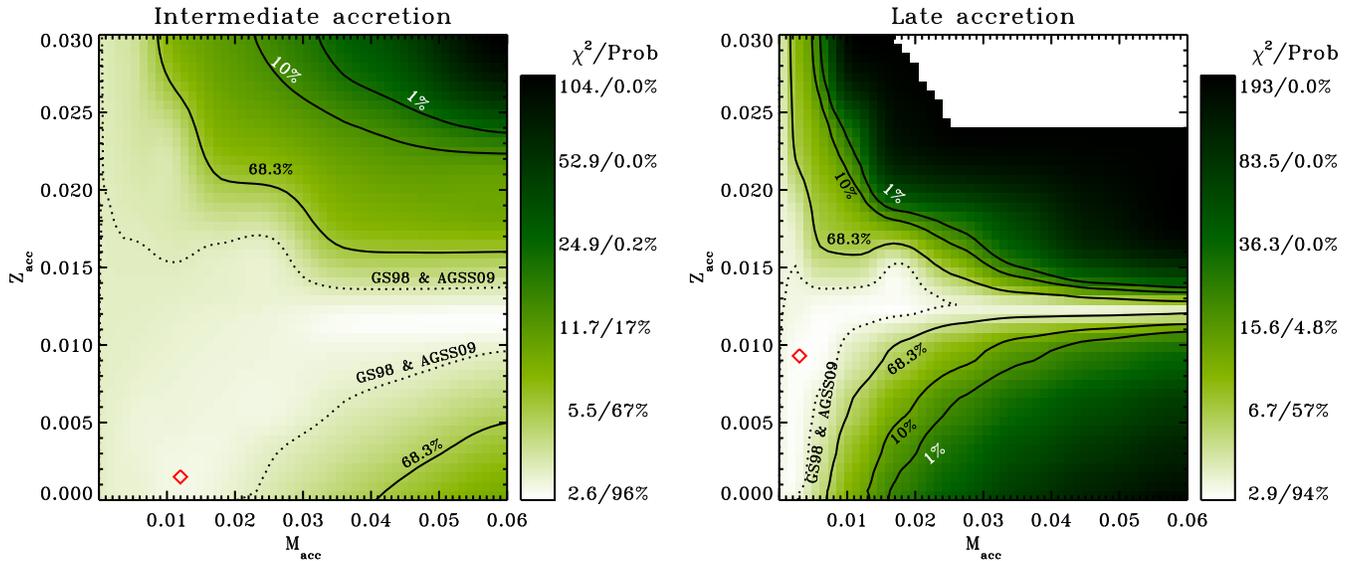}
\caption{Results for the combined  $\chi^2$ global solar neutrino analysis for
  the intermediate  and late  accretion scenarios. Equal  probability contours
  are shown  for 1\%, 10\% and 68.3\%.   The best fit models  are indicated by
  red diamonds.   The fixed-$\chi^2$ contours corresponding to  the AGSS09 and
  GS98 SSM fits (Table~\ref{tab:neutrinos})  are overlaid, to indicate regions
  where accretion models do better or worse than the SSMs.
\label{fig:chi2}
}
\end{figure*}

Among  the four  well-determined  fluxes,  the $^8$B  neutrinos  are the  most
sensitive  to  accretion,  responding   to  the  variations  in  initial  core
metallicity  that  occur because  the  accretion  model  is solved  under  the
constraint  of a  fixed final  convective-zone  metallicity-to-hydrogen ratio.
The $^8$B solar  flux is also the most certain, now  determined to $\sim 3\%$.
Moreover,  $\Phi(^8$B) provides  information complementary  to helioseismology
because  the  production region,  $R  \lesssim  0.1\,  \rsun$, is  essentially
inaccessible  to  p-mode  study.   In  fact,  we find  that  all  models  with
metal-rich accretion, because  of their lower $\zini$ values,  produce too few
$^8$B    neutrinos.    Results    are   shown    in   panel    {\em    a}   of
Figures~\ref{fig:nuv0a5t10},~\ref{fig:nuv0a15t10},                          and
\ref{fig:nuv0a30t10}. This  suggests that metal-rich accretion  would create a
conflict  smaller than  but reminiscent  of  the old  solar neutrino  problem,
particularly for the intermediate  and late accretion scenarios.  In contrast,
models with  metal-poor accretion,  for an appropriate  choice of  $\mac$ that
depends  on the  accretion scenario  (early,  intermediate, or  late), can  be
brought  into excellent  agreement  with the  experimental  fluxes.  The  high
sensitivity of the $^8$B flux to  metallicity leads to $\mac$ values as low as
0.01~$\msun$  being  disfavored  for  the  extreme  case  of  late  metal-free
accretion.

The $^7$Be, pp, and pep fluxes, are now also determined to a high level of
precision,   but   are   less   sensitive   to  the   core   conditions   (see
Figures~\ref{fig:nuv0a5t10},~\ref{fig:nuv0a15t10}, 
  and  \ref{fig:nuv0a30t10}).  With  respect  to the  AGSS09  SSM, a  moderate
  amount of  metal-poor accretion can lead  to improvements in  both the $^8$B
  and $^7$Be fluxes, as discussed below. A slightly more metal-rich core seems
  preferred by these fluxes. 
 
We have  done a global  analysis of neutrino  fluxes, with the results  of the
$\chi^2$      calculations     given     in      panel     {\em      f}     of
Figures~\ref{fig:nuv0a5t10},~\ref{fig:nuv0a15t10},  and  \ref{fig:nuv0a30t10}.
In the case of early accretion, only the most 
metal-rich models  with massive  accretion show significant  disagreement with
solar data.   The neutrino flux responses  to accretion are  more sensitive in
the intermediate and late scenarios, however.  Figure~\ref{fig:chi2} gives the
resulting $\chi^2$  surfaces in the $\mac  - \zac$ plane  for these scenarios.
In each  panel the  best-fit model is  denoted by  a red diamond.  Solid lines
depict the  68.3\%, 10\%, and 1\%  probability contours, and  dotted lines the
contours of probability for the AGSS09 and GS98 SSMs (90\%) 
(Table~\ref{tab:neutrinos}).  In the late accretion case, the top-right corner
is  empty  because  no  solar   models  satisfying  our  requirements  can  be
constructed  for such $\mac  -\zac$ combinations.   These plots  define rather
clearly the portions of the $\mac  - \zac$ parameter space that are consistent
with  solar neutrino  data.   For intermediate  accretion,  all $\mac$  values
considered for $\zac \lesssim 0.015$ yield excellent fits, comparable to SSM
results. Improvements are possible but minor because, as we have
noted before, AGSS09 SSM neutrino flux predictions are in very nice agreement
with current  solar neutrino data.  We  conclude that solar  neutrino data are
primarily 
of value in  ruling out accretion scenarios that  give unacceptable fluxes (as
opposed to providing positive evidence for accretion scenarios that marginally
improve  neutrino flux  fits).  As  the accretion  becomes more  metal-rich, a
larger range of  high-$\mac$ accretion scenarios is excluded.   In the case of
late accretion,  large fractions of the $\mac$-$\zac$  plane are significantly
disfavored  by solar  neutrino  data  at both  the  metal-rich and  metal-poor
extremes.  The  full $\mac$  range is allowed  only for accretion  models with
$\zac \approx 0.015$: these are cases close to the SSM.

\begin{figure}[t]
\includegraphics[scale=.55]{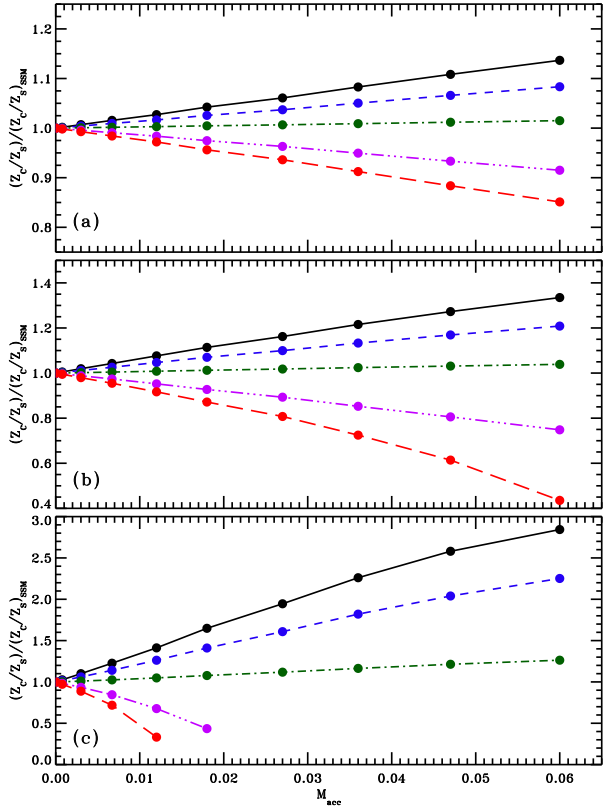}
\caption{Ratio   of   the  central   to   surface   metallicity  in   selected
  models. Panels  {\em a}, {\em b},  and {\em c} refer  to early, intermediate
  and late accretion scenarios  respectively.  Metallicity sequences are shown
  for $\zac=0, \,  0.0059, \, 0.0130, \, 0.0240, \, 0.0300$  (top to bottom in
  each panel).  
\label{fig:zczs}
}
\end{figure}

We  close this  section on  solar  neutrinos with  a short  discussion of  the
potential role of neutrino fluxes from the CN cycle.  These fluxes
are very sensitive  to metallicity because they respond in  two ways: like the
$^8$B  neutrinos, the  CN  neutrinos respond  sharply  to {\em  environmental}
effects associated with metallicity --  changes in the radiative opacities and
the mean molecular  weight that alter the temperature of  the solar core.  But
the CN  fluxes have an additional  linear dependence on the  core abundance of
the CN elements, as C and  N catalyze the hydrogen-burning reactions of the CN
I  cycle.   \citet{haxton:2008}  have   shown  that  a  future  {\em  precise}
measurement of  $\Phi(^{13}$N) and $\Phi(^{15}$O)  could be used  to determine
the solar  core C+N abundance  to $\sim$ 10\%,  including all solar  model and
neutrino  parameter   uncertainties  (see  also   \citealt{PGSH}).   A  direct
determination  of the  central  C+N abundance  would  be very  valuable, as  a
comparison to  surface abundances  would constitute a  direct test of  the SSM
hypothesis  of  a homogeneous  ZAMS  Sun.   In our  models  we  find that  the
central-to-surface metallicity  ratio, normalized to the SSM  value to account
for the  effects of  microscopic diffusion, deviates  from unity by  more than
10\%  in  many of  our  models, particularly  for  the  intermediate and  late
accretion  scenarios,   as  illustrated  in   Figure~\ref{fig:zczs}.   Thus  a
high-quality CN  flux measurement could help  rule out a large  portion of the
potential  parameter space  of  accretion models,  making  the measurement  an
important diagnostics of the early Sun and its proto-planetary disk.
Of course,  this is  in the  context of the  assumptions we  have made  in our
accretion scenarios,  which include  a uniform rate  of accretion  of material
with  a  metallicity related  to  AGSS09  by a  simple  scale  factor, onto  a
proto-Sun  with  a  metallicity  that  is  also scaled  to  AGSS09.   As  disk
condensation  temperatures  vary  from  element  to element,  there  is  ample
motivation for  relaxing these assumptions.  (More general  scenarios in which
such assumptions are reconsidered  will be discussed elsewhere.)  Despite this
caveat,  we  believe  that  the  $^{13}$N and  $^{15}$O  neutrino  fluxes  are
potentially very  useful as probes  of solar core metallicity  and, therefore,
potentially of the chemistry of the proto-solar disk.  This chemistry sets the
conditions for the formation of the solar system's diverse planets, and, as we
have speculated, may  have also altered the composition  of the solar surface.
Given  that there exist  very few  quantitative probes  of the  processes that
governed solar system formation, it is important to pursue any experiment that
can directly constrain phenomena like accretion during the formation period.

\subsection{Comparison with previous works}
\label{sec:cmpothers}

\citet{guzik:2006,guzik:2010}  and \citet{castro:2007} have  studied accretion
of metal-poor material as a  possible solution to the solar abundance problem.
The authors constructed solar models in which the interior metallicity is high
(around the GS98  value) and the envelope metallicity  is low, consistent with
\citet{ags05}.   \citet{guzik:2006} assumes  that  a mass  of 0.02~$\msun$  is
accreted  during   a  period  of  36~Myr,   onto  a  ZAMS   stellar  model  of
0.98~$\msun$.  \citet{castro:2007} takes  a simplified  approach in  which the
metallicity of the convective envelope is simply scaled down by a given factor
(0.5) when the solar model is 74~Myr old.  In both treatments the accretion is
assumed  to occur after  the contraction  phase to  the MS  is over,  when the
convective envelope is  thin and very close to  its present-day extension.  In
this aspect, their models resemble our late accretion scenario.

\begin{figure}[t]
\includegraphics[scale=.55]{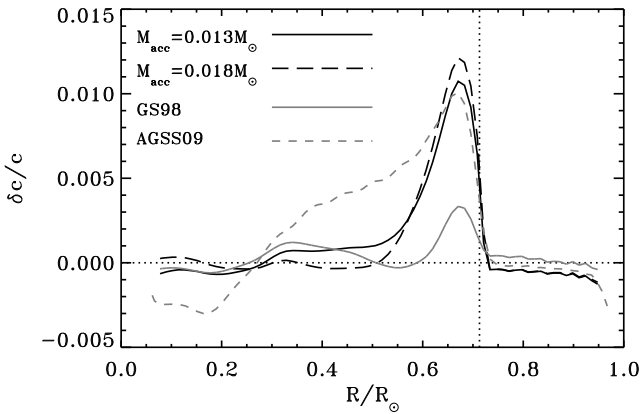}
\caption{Sound speed profiles for the models with $\mac=0.013$ (black solid
  line) and $0.018~\msun$ (red dashed-dotted line),  $\zac=0.0093$,
  $\taui=30$~Myr,  and $\Delta \tau_{\rm   ac}=10$~Myr. GS98 (grey solid
  dashed) and    AGSS09 (grey dashed line)  standard solar models are shown
  for comparison.  
\label{fig:dccomp}
}
\end{figure}

In terms  of $\zini$,  two of our  late-accretion models closely  resemble the
cases  studied  \citet{guzik:2006}  and  \citet{castro:2007}.  These  are  our
models   with   $\zac=0.0093$  and   with   $\mac=0.013$  and   $0.018~\msun$,
respectively  (see Figures~\ref{fig:hv0a30t10}~and~\ref{fig:nuv0a30t10}).  The
sound speed  profiles of these  two models, shown  in Figure~\ref{fig:dccomp},
match very well that of the GS98 SSM below $R/R_\odot=0.5$, but the bump below
the convective zone remains quite prominent. Therefore $\dc$ shows only modest
improvement over the AGSS09 SSM  value.  These results correspond to Figure 12
of \citet{guzik:2006} and to Figure 1 of \citet{castro:2007}.  Our models also
yield $\rcz= 0.725$  and $0.727\, \rsun$, respectively, so  the agreement with
helioseismology in  this quantity  is slightly worse  than for the  AGSS09 SSM
($0.723\,  \rsun$).   These observations  are  in  qualitative agreement  with
\citet{castro:2007},  who also find  that the  convective zone  gets shallower
(0.732~$\rsun$  instead  of  0.730~$\rsun$)   in  the  model  with  accretion.
\citet{guzik:2006},  on  the other  hand,  finds  improvement  in $\rcz$  with
respect  to the  model with  no accretion.  The reason  for  the qualitatively
different  behavior is  not clear,  but plausibly  could be  connected  to the
assumed effects of the accretion  on the convective zone's hydrogen and helium
abundances.  Finally, our models give $\ys=0.244$ and 0.250, in good agreement
with    the    helioseismic     value,    whereas    \citet{guzik:2006}    and
\citet{castro:2007} find  $\ys=0.240$, marginally compatible  with the seismic
result. The  small numerical discrepancies may  be due to the  use of somewhat
different solar compositions in the  three sets of calculations.  Yet, overall
the resulting models are in reasonable agreement, given the differences in the
manner accretion was implemented in the three studies.

\section{Summary and Conclusions}\label{sec:conclusions}
 
In  this study  we  have explored  the  possibility that  an accretion  phase,
occurring after the Sun developed a radiative core, could resolve, or at least
mitigate, the  solar abundance problem.   The motivation for  considering this
possibility is the observation  that large-scale metal segregation occurred in
the  proto-planetary  disk,  connected  with  the formation  of  the  planets.
Extending  previous  studies, we  considered  both  metal-poor and  metal-rich
accretion and also allowed accretion to occur at different evolutionary stages
of  the young  Sun. We  also considered  the constraints  on models  from both
helioseismology and  neutrino measurements,  while earlier studies  focused on
the former.

We began  by updating the  SSM, adopting the  nuclear reaction rates  from the
very recent evaluation  of SFII.  The new SSMs for  ASSS09 and GS98 abundances
then  were  used  as  benchmark  calculations against  which  the  effects  of
accretion could be measured.  The  new models differ only modestly from others
recently  calculated  \citep{ssm:09}. The  new  solar  models  with the  lower
metallicity AGSS09  composition remain in conflict  with helioseismology.  The
newly recommended nuclear fusion rates  do impact predicted SSM solar neutrino
fluxes.  The higher  metallicity GS98 SSM remains in  excellent agreement with
solar  neutrino  data, but  now  the lower  metallicity  AGSS09  SSM shows  an
equivalent level of agreement. This also  partly the result of the most recent
$^7$Be flux measurement by Borexino \citep{borex:2011}. Thus solar neutrino
spectroscopy (and  the associated laboratory  astrophysics) has not  yet reach
the level  of precision necessary to  distinguish between the  AGSS09 and GS98
SSMs.

The main conclusion we draw from our study of accretion is that, for the class
of  models  we explored,  no  scenario was  found  to  provide a  satisfactory
solution  of the solar  abundance problem.   Here ``satisfactory  solution" is
defined as a  model with AGSS09 surface abundances that  agrees with the known
helioseismic  properties as well  as the  GS98 SSM.   We find  that metal-rich
accretion can bring  the depth of the convective  envelope into agreement with
the seismic  value, but that the  resulting surface helium  abundances in such
models are  always too low  -- in  fact, in some  of these models  the initial
helium  abundance falls  below the  primordial SBBN  value.   Conversely, some
metal-poor accretion  models bring surface  helium abundance into a  very nice
agreement with the solar value, but  at the cost of a convective envelope that
is even shallower than that of the AGSS09 SSM.

Problems in the sound speed profile are generally strongly correlated with the
problems  in the depth  of the  convective envelope.   The better  (worse) the
latter, the better  (worse) the former.  An exception to  this general rule is
found for  late accretion  models in which  about 0.015~$\msun$  of metal-poor
material is accreted.  In this case, though the model's convective envelope is
too shallow, there is some improvement in the sound speed profile.  However, a
significant bump  in the sound  speed difference (model  vs.  heliosesimology)
remains in the region right  below the convective envelope.  These models have
surface helium abundances and neutrino fluxes that agree with solar values.

Current  solar neutrino  data  constrain solar  core  properties.  Our  global
comparison of model and solar  fluxes rules out both metal-poor and metal-rich
accretion for $\mac \gtrsim 0.01\, \msun$ in the late accretion scenario.
 
The results of our accretion study provide further motivation for the proposal
to exploit future measurements of CN  neutrinos as a test of the C+N abundance
of the  solar core  \citep{haxton:2008}.  We find  that the ratio  between the
surface and central solar metallicity  in accretion models deviates from unity
(gravitational  settling  taken into  account)  by  more  than 10\%  for  most
accretion scenarios.  Such  a measurement of the solar  core bulk abundance of
C+N does not directly constrain other metals and thus the overall metallicity:
despite the simplifying  assumptions made in our accretion  model, there is no
{\it  a priori}  reason to  believe that  the composition  of accreted  gas is
related  to  the  primordial  gas  by  a  simple  scale  factor.   Still,  any
quantitative abundance  determination in the  solar interior would be  a major
step forward.

In  the past, accretion  of metal-rich  material was  discussed as  a possible
solution to the solar neutrino problem \citep{winnick:2002}.  More recently, a
new  low-metallicity solar composition  has generated  concern that  the Sun's
surface  might  have  been  altered  in an  episode  of  metal-poor  accretion
\citep{guzik:2006,castro:2007}.   Additional  motivations  for continuing  the
study of  accretion onto the  early Sun can  be found in suggestions  that the
mechanism generating metal-poor or  metal-rich accretion might be connected to
the  dynamics  of  the  proto-planetary  disk.   The  possibility  of  such  a
connection has  been raised in several  papers: in the  observation that solar
abundance problem  involves a deficit  of convective-zone metal  comparable to
the excess of metal sequestered  in the planets \citep{haxton:2008}; in recent
models of the  assembly of low mass stars  \citep{baraffe:2010}; and in claims
that the  Sun has an  abundance pattern of  metals that correlates  with metal
condensation temperatures, suggesting a connection with disk chemistry and the
formation  of  rocky  planets  \citep{twins1}.  Work  on  accretion  scenarios
motivated by these observations is ongoing and will be reported elsewhere.

\acknowledgements

Aldo  Serenelli is  partially supported  by the  European  Union International
Reintegration Grant  PIRG-GA-2009-247732, the MICINN  grant AYA08-1839/ESP, by
the  ESF EUROCORES Program  EuroGENESIS (MICINN  grant EUI2009-04170),  by SGR
grants  of   the  Generalitat  de   Catalunya  and  by  the   EU-FEDER  funds.
W. C. Haxton is supported in  part by the U.S. Department of Energy, including
under   DE-SC00046548  at   Berkeley.   CPG   is  supported   by   the  grants
FPA-2007-60323 and PROMETEO/2009/116. We thank Maria Bergemann for useful
discussions and her careful reading of the manuscript.  

\bibliographystyle{apj}

\end{document}